\documentclass[traditabstract]{aa}
\usepackage{graphicx}
\usepackage{txfonts}
\usepackage{longtable}
\usepackage[authoryear]{natbib}
\bibpunct{(}{)}{;}{a}{}{,}

\def\ltsima{$\; \buildrel < \over \sim \;$}
\def\simlt{\lower.5ex\hbox{\ltsima}}
\def\gtsima{$\; \buildrel > \over \sim \;$}
\def\simgt{\lower.5ex\hbox{\gtsima}}
\newcommand{\HI}{\ion{H}{i}}

\begin{document}
   \title{The extended structure of the dwarf irregular galaxies Sextans~A and Sextans~B}
\subtitle{Signatures of tidal distortion in the outskirts of the Local Group\thanks{Based on data acquired using the Large Binocular Telescope (LBT). The LBT is an international collaboration among institutions in the United States, Italy, and Germany. LBT Corporation partners are The University of Arizona on behalf of the Arizona university system; Istituto Nazionale di Astrofisica, Italy; LBT Beteiligungsgesellschaft, Germany, representing the Max-Planck Society, the Astrophysical Institute Potsdam, and Heidelberg University; The Ohio State University; and The Research Corporation, on behalf of The University of Notre Dame, University of Minnesota and University of Virginia.} }

   \author{M. Bellazzini\inst{1}, G. Beccari\inst{2}, F. Fraternali\inst{3,5}, 
           T.A. Oosterloo\inst{4,5},  A. Sollima\inst{1}, 
           V. Testa\inst{6}, S. Galleti\inst{1},  
           S. Perina\inst{7}, M. Faccini\inst{6} \and F. Cusano\inst{1}
           }
         
      \offprints{M. Bellazzini}

   \institute{INAF - Osservatorio Astronomico di Bologna,
              Via Ranzani 1, 40127 Bologna, Italy\\
            \email{michele.bellazzini@oabo.inaf.it} 
            \and
            European Southern Observatory, Av. Alonso de Cordova, 3107, 19001 Casilla, Santiago, Chile
            \and
            Dipartimento di Astronomia - Universit\`a degli Studi di Bologna,
            Via Ranzani 1, 40127 Bologna, Italy
            \and
            Netherlands Institute for Radio Astronomy, Postbus 2, 7990 AA Dwingeloo, the Netherlands
            \and
            Kapteyn Astronomical Institute, University of Groningen, Postbus 800, 9700 AV Groningen, 
            The Netherlands
            \and
             INAF - Osservatorio Astronomico di Roma, via Frascati 33, 00040 Monteporzio, Italy
            \and
            Institute of Astrophysics, Pontificia Universidad Cat—lica de Chile, Avenida Vicu\~na Mackenna 4860, Macul, Santiago, Chile 
           }

     \authorrunning{M. Bellazzini et al.}
   \titlerunning{The extended structure of the dwarf irregular galaxies Sextans~A and Sextans~B}

   \date{Accepted for publication by A\&A }

\abstract{
We present a detailed study of the stellar and \HI\ structure of the dwarf irregular galaxies Sextans~A and Sextans~B, members of the NGC~3109 association. We use newly obtained deep (r$\simeq 26.5$) and wide field g,r photometry to extend the Surface Brightness (SB) profiles of the two galaxies down to $\mu_V\simeq 31.0$~mag/arcsec$^2$. We find that both galaxies are significantly more extended than what previously traced with surface photometry, out to $\sim 4$~kpc from their centers along their major axis. Older stars are found to have more extended distribution with respect to younger populations. We obtain the first estimate of the mean metallicity for the old stars in Sex~B, from the color distribution of the Red Giant Branch, $\langle [Fe/H]\rangle=-1.6$. The SB profiles show significant changes of slope and cannot be fitted with a single Sersic model. Both galaxies have HI~ discs as massive as their respective stellar components. In both cases the \HI~ discs display solid-body rotation with maximum amplitude of $\sim 50$~km~s$^{-1}$ (albeit with significant uncertainty due to the poorly constrained inclination), implying a dynamical mass $\sim 10^{9}~M_{\sun}$, a mass-to-light ratio $\frac{M}{L_V}\sim 25$ and a dark-to-barionic mass ratio of $\sim 10$. The distribution of the stellar components is more extended than the gaseous disc in both galaxies. We find that the main, approximately round-shaped, stellar body of Sex~A is surrounded by an elongated low-SB stellar halo that can be interpreted as a tidal tail, similar to that found in another member of the same association (Antlia).
We discuss these, as well as other evidences of tidal disturbance, in the framework of a past passage of the NGC~3109 association close to the Milky Way, that has been hypothesized by several authors and is also supported by the recently discovered filamentary configuration of the association itself.}


\maketitle
%

\section{Introduction}
\label{intro}

The origin of the variety of morphologies of dwarf galaxies and the evolutionary path leading to the formation of the amorphous and gas-devoid dwarf Spheroidal (dSph) galaxies, are generally recognized as key questions of modern astrophysics, having important consequences for our understanding of galaxy formation and cosmology \citep{mateo,tht}. The local morphology-density relation (i.e. dSph are found in the surroundings of giant galaxies while gas-rich dwarf Irregulars - dIrr - are more far away, on average) lead to the idea that strong interactions with the large galaxies they are orbiting around transformed primordial actively star-forming dIrrs into present-day quiescent dSph, removing the gas and stopping the star formation \citep{mateo}. 
Recent state-of-the-art modeling in cosmological context revealed that tidal interactions and ram-pressure stripping can be quite efficient in transforming gas-rich disc dwarfs (plunging deeply into the halo of their main galaxy) into dSphs \citep[the tidal stirring model,][]{lucionat}, but also SN winds and the cosmic UV background are found to play a significant roöle in removing gas from dwarfs \citep{kaza,saw10,saw12}. 

Observations of isolated\footnote{Here the terms {\em isolated} means ``several hundreds of kpc distant from a giant galaxy", specifically the Milky Way and M31, since we focus our research to the Local Group and its surroundings, where dwarfs can be resolved into individual stars. In practice we selected our targets also to lie more than 200~kpc away from other dwarfs. It is obvious that finding a galaxy far away from other LG members at the present epoch does not guarantee that it has evolved in isolation \citep[see, e.g.,][]{teys12}.} dwarfs in the Local Group (LG) is expected to provide a crucial insight on the impact of the various factors in shaping the present-day status of dSphs. Isolated dwarfs may represent the test case of evolution without interactions and can provide constraints on the initial conditions at the epoch of formation. These considerations were the driving case for a large Hubble Space Telescope (HST) programme \citep{lcid} that performs extremely deep observation in small central fields of a bunch of isolated galaxies, searching for the effects of SN feedback and/or cosmic re-ionization, as they may be recorded in the Star Formation History (SFH) of the systems \citep[see, e.g.][]{monel_cet,monel_tuc}. 

We have recently started a research project aimed at exploring a fully complementary aspect, i.e. the large scale structure of these galaxies, considering both their stellar bodies and of their neutral gas components (\HI). The lack of tidal limits imposed by nearby masses implies that isolated galaxies may preserve extended stellar haloes at low SB, as it has been indeed observed in some  cases (e.g., \citet{vanse} in Leo A, \citet{sanna} in IC10). These feeble but extended structures record crucial information on the formation and evolution of these galaxies, and provide the precious stellar test-particles to study the dynamics of these systems at large distances from their centers, thus probing their Dark Matter (DM) halos over a wide radial range. The structure and kinematics of the neutral Hydrogen provide additional and complementary information. It must be stressed that dwarf galaxies evolved in isolation should have preserved their pristine DM haloes virtually untouched since the beginning of time, hence bearing fundamental information on their original structure and mass distribution.

In a pilot project focused on the extremely isolated dwarf VV124=UGC4879 \citep[][hereafter Pap-I]{pap1} we demonstrated that the outer structures of these galaxies can be traced down to extremely low Surface Brightness (SB) levels ($\mu_V\simeq 30$ mag/arcsec$^2$) with star counts from very deep photometry obtained with state of the art wide-field cameras on 8m class telescopes, with modest amounts of observing time, under excellent seeing conditions. In that specific case we found that (a) the stellar body of the galaxy was much more extended than previously believed (from $\sim 500$~ to $\ga 1500$~pc), and (b) that the galaxy was highly structured in its outskirts, showing two extended and thin wings whose origin is still to be clarified \citep[see Pap-I and][]{kvo124,kv124}. 

In this paper we present the results of the same kind of analysis performed for VV124 in Pap-I \citep[and followed up also with Hubble Space Telescope (HST) photometry in][]{vvhst}, for two additional targets in our survey of nearby isolated dwarfs, the dwarf Irregular (dIrr) galaxies Sextans~A (DDO~75) and Sextans~B (DDO~70).
Both galaxies contain a few $10^7$ solar masses of HI \citep{ott} and are actively forming stars at the rate of 2-8$\times 10^{-3}~M_{\sun}~yr^{-1}$ \citep{weisz}.
They are members of a very loose group of dwarf galaxies \citep[the NGC~3109 association, see][]{tully06} that is located in the outskirts of the LG, more than 1~Mpc away from the Milky Way (MW) and M31.
Triggered by the results shown in Sect.~\ref{struc}, below, we have recently noted that all the known members of the association plus the newly discovered Leo~P dwarf \citep{giova} are strictly clustered 
around a line in space and display a tight velocity gradient with the distance along this line 
\citep{filam}. This can indicate that the members of this group may have been subject to a significant tidal interaction in the past. 

In any case, Sex~A and Sex~B are separated by more than 250~kpc and  
they are $\sim 500$~kpc and $\sim 700$~kpc apart from the most massive member of the association (NGC~3109), hence they appear as remarkably isolated galaxies, at the present epoch.  For this reason they were fully eligible as targets for our survey. It is clear that one of the possible outcomes of our analyses is to find out that some of our target did not evolve in isolation in the past, as it now seems to be the case for Sex~A, at least. This may open an interesting window on the role of interactions in low density groups of low mass galaxies.

The plan of the paper is the following: in Sect.~\ref{phot} we describe the optical photometry observations and their data reductions and calibrations, in Sect.~\ref{grad} we provide constraints on the metallicity distribution of old stars, and we briefly consider the spatial distribution of the various stellar species within the two galaxies, an analysis that was never performed before on such wide scales. In Sect.~\ref{struc} we present the SB profiles and density maps: also in these cases we are able to trace the stellar structure of the galaxies out to distances from their centers never explored before. In Sect.~\ref{HI} we re-consider the structure and dynamics of the \HI\ distributions associated to Sex~A and Sex~B and, finally, in Sect.~\ref{disc} we summarize and discuss our results. Moreover there are two appendices:
in Appendix~\ref{quality} we present a direct comparison between our own photometry and HST photometry for Sex~A and Sex~B, while in Appendix~\ref{clus} we report on the search for compact star clusters we performed on our images.

\section{LBT Observations and data reduction}
\label{phot}

\begin{figure}
   \centering
   \includegraphics[width=\columnwidth]{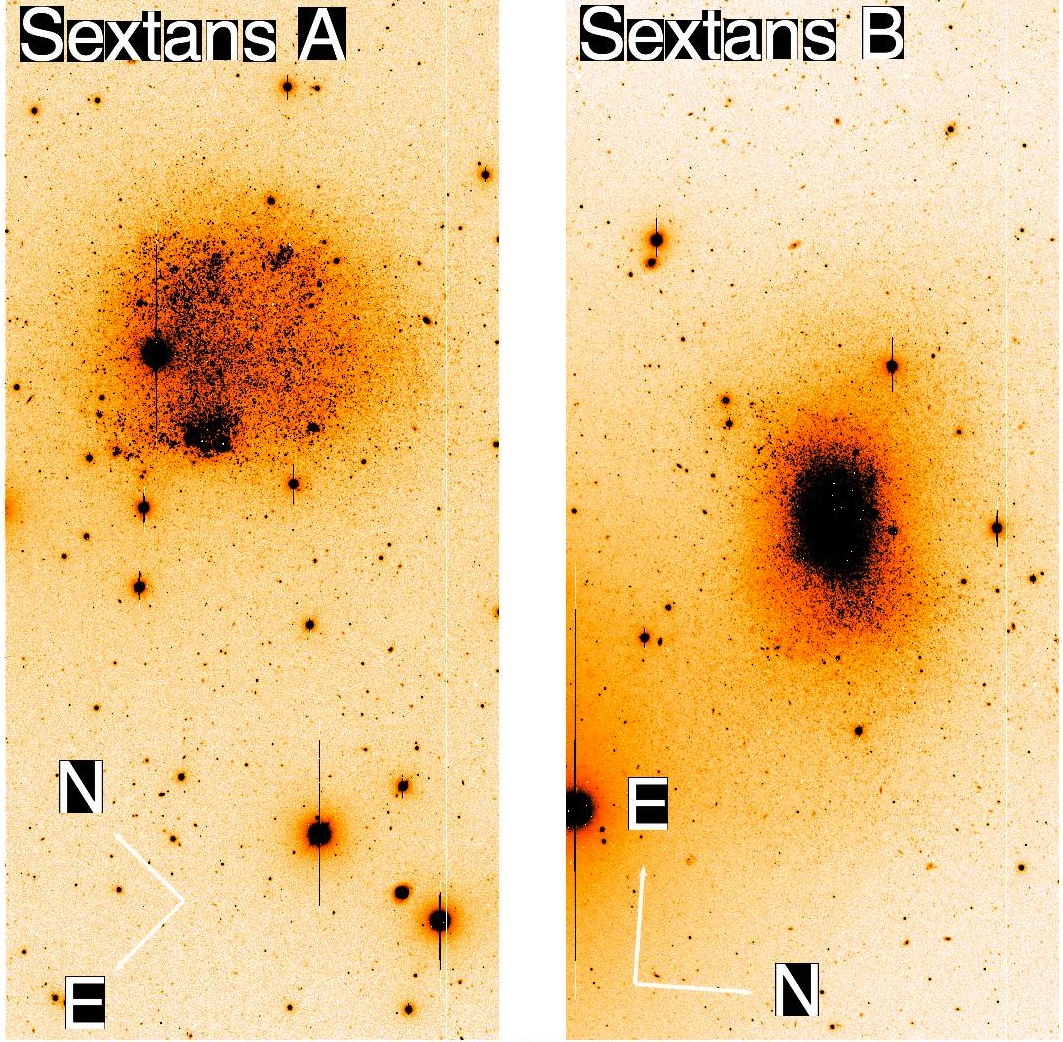}
     \caption{Deep g band images of a large portion of Chip~2 for Sex~A (left panel) and Sex~B (right panel). 
     The sampled FoV is $7.7\arcmin\times16.2\arcmin$}
        \label{imaC}
    \end{figure}

Deep $g$ and $r$ photometry was acquired on the night of February 21, 2012, at LBT, using the Large Binocular Camera \citep[LBC, ][]{lbc} in binocular mode; $g$ images were acquired with the blue arm and $r$ images with the red arm of the telescope/camera.
The optics of each LBC feed a mosaic of four 4608~px~$\times$~2048~px CCDs, with a pixel scale of 0.225 arcsec~px$^{-1}$. Each CCD chip covers a field of $17.3\arcmin\times7.7\arcmin$. Chips 1, 2, and 3  are flanking one another, being adjacent along their long sides; Chip 4 is placed perpendicular to this array, with its long side adjacent to the short sides of the other chips \citep[see Fig.~4 of][]{lbc}. During our observations the pointing was chosen to place the targets near the center of Chip~2, with the long side nearly aligned with the major axis of the galaxy (see Fig.~\ref{imaC}). In the following, we will use the terms Chip~1(2,3,4) and field~1(2,3,4), abbreviated as f1, f2, f3 and f4, interchangeably.
Five long ($t_{exp} = 300$~s) and two short ($t_{exp} = 20$~s) exposures per filter were acquired during the night; the seeing ranged from $0.70\arcsec$ to $1.2\arcsec$. The average seeing and background level were worse, on average, for the Sex~A set of images , resulting in a slightly lower limiting magnitude with respect to the photometry of Sex~B\footnote{In particular for the deep r images, with a mean seeing of $1.08\arcsec$ and $0.86\arcsec$ for Sex~A and Sex~B, respectively.}. 
Moreover, Sex~B images were obtained at airmass$\simeq 1.13$ while those of Sex~A have airmass$\simeq 1.27$
The short exposures were obtained to provide a bridge between the photometry from our long exposures (reaching the saturation level at $r\simeq 18.0$) and the secondary calibrators from the SDSS DR9 catalog \citep{dr9} that can be as faint as $g,r\simeq 23.0$ but have average photometric errors $\sigma_g,\sigma_r\le0.03$ mag only for $g,r\le 20.0$.

Relative photometry was performed independently on each long exposure image using the Point Spread Function (PSF)-fitting code DAOPHOTII/ALLSTAR \citep{daophot,allframe}. Sources with peaks higher than 3$\sigma$ above the background were identified in a stacked image obtained by registering and co-adding all the images considered for the analysis. Then, each of these stars was re-identified and fitted on each image (when possible). Only sources found at least in three $g$ {\em and} three $r$ images were retained in the final catalogue. The average and the standard error of the mean of the independent measures obtained from the different images were adopted as the final values of the instrumental magnitude and of the uncertainty on the relative photometry. To clean the catalog from spurious sources, we also adopted cuts in the image quality parameters $CHI$ and $SHARP$, provided by DAOPHOTII. After accurate inspection of the distribution of measured sources in the planes mag vs. $CHI$ and mag vs. $SHARP$, we decided to retain only sources having $CHI \le 2.0$ and SHARP within magnitude-dependent contours traced to include
the bulk of stellar objects. Considering all the four fields together, this selection removed 12745 (15864) sources from the Sex~A (Sex~B) original dataset, leaving a final catalog of 52598 (68369) retained sources. Aperture photometry (corrected to infinite apertures) was independently obtained for short exposures images using Sextractor \citep{sex}.

\subsection{Astrometry and Photometric Calibration}

\subsubsection{Sextans~B}
\label{sexb}

The instrumental positions were transformed (independently for each chip) into J2000 celestial coordinates using the stars in common with the SDSS DR9 catalog \citep{dr9}. The adopted astrometric solutions, in the form of third degree polynomials, were obtained with CataXcorr\footnote{CataXcorr is a code aimed at cross-correlating catalogues and finding astrometric solutions, developed by P. Montegriffo at INAF - Osservatorio Astronomico di Bologna, and successfully used by our group during the past 10 years.} using more than 100 stars per field; the r.m.s. scatter of the solutions  was $\le 0.1\arcsec$ in both RA and Dec. 

Also the instrumental magnitudes ($g_i$, $r_i$) were transformed into the SDSS ugriz absolute photometric system using stars in common with SDSS DR9, with a first order polynomial as in Pap-I. In particular, photometry from short exposures was calibrated first and then the solution was transferred to the main catalog obtained from deep images. The rms scatter about the photometric solutions is $\simeq 0.04$~mag, implying an error on the zero point of $\simeq 0.01$ mag.  
In the following analysis we will use equations from 3 to 6 of Pap-I to transform $g$,$r$ magnitudes into V,I, when needed. 

\subsubsection{Sextans~A}
\label{sexa}

Sex~A is not included into the SDSS footprint, hence photometric secondary standards are lacking in the field. Since the observations were acquired within a few hours from those of Sex~B we adopted the same photometric solution, corrected for the difference in airmass between the two sets of observations. Clearly the absolute photometric calibration of the Sex~A data is more uncertain with respect to the case of Sex~B. Still, a direct comparison with the HST photometry publicly provided by the ANGST team \citep{dal09}\footnote{\tt http://www.nearbygalaxies.org/} shows that the zero points of the two datasets are in satisfactory agreement
(within $\sim 0.03$~mag for Sex~B and within $\sim 0.10$~mag for Sex~A). We note that a very accurate photometric calibration for this kind of data is both (a) difficult to achieve (see Pap-I, for discussion) and, (b) not crucial for the scientific goals of the present survey. 
The astrometric solution (analogous to that described above for Sex~B) was derived from GSC2.2 standard stars, using more than 100 stars per field; the r.m.s. scatter of the solutions  was $\le 0.3\arcsec$ in both RA and Dec. 
A final cross check with the catalog by \citet{massey} revealed a r.m.s. scatter $\le 0.15\arcsec$ in both RA and Dec, for the stars in common with this dataset.

\subsection{Reddening and Distance}

We interpolated the \citet{ebv} reddening maps - as re-calibrated by \citet{schlaf} - to obtain an estimate of $E(B-V)$ and its variation over the considered FoVs. We adopted a regular grid with knots spaced by $3.0\arcmin$, and found that (a) the mean reddening values are slightly lower than those usually reported in the literature \citep[see, e.g.][hereafter M12]{dal09,mcc}, due to Schlafy et al.'s re-calibration, and (b) the reddening variations are negligible (see Table~\ref{Tab_parA} and \ref{Tab_parB}). In the following, as in Pap-I, we adopt the reddening laws $A_g=3.64E(B-V)$ and $A_r=2.71E(B-V)$, derived by \citet{gira} for cool metal-poor giants. We note that the choice of the reddening law is not particularly relevant in these cases, given the very low extinction regime of the considered fields.

To obtain distances fully consistent with the scale adopted in Pap-I we used the accurate V,I HST photometry from ANGST to (a) correct the ANGST CMD for the newly derived values of the reddening, and (b) re-derive the magnitude and mean colors of the RGB-Tip as done in Pap-I. We obtain $I_{0,tip}= 21.72 \pm 0.06$ and $(V-I)_{0,tip}=1.39$ for Sex~A, and $I_{0,tip}= 21.75 \pm 0.06$ and $(V-I)_{0,tip}=1.41$ for Sex~B, in excellent agreement with \citet{dal09}. Using the following equation
for the absolute magnitude of the RGB Tip:

\begin{equation}
M_I^{TRGB} = 0.080(V-I)_0^2 -0.194(V-I)_0 -3.939
\end{equation}

from \citet{cefatip}\footnote{Derived from the original calibrations as a function of [Fe/H] and [M/H] presented in \citet{tip1} and \citet{tip2}.}, we obtain $M_I^{TRGB} = -4.05\pm 0.10$, for both galaxies. Finally, we obtain $(m-M)_0 = 25.77 \pm 0.12$ for Sex~A, and $(m-M)_0 = 25.80 \pm 0.12$ 
for Sex~B, corresponding to $D=1.42\pm0.08$~Mpc and $D=1.44\pm0.08$~Mpc, respectively, where the reported errors include {\em all the statistic and systematic sources of error}. The derived distance moduli are slightly larger than those by \citet{dal09}, because of the small differences in the adopted reddening and $M_I^{TRGB}$, still they are fully compatible within the uncertainties.

\subsection{The Color Magnitude diagrams}

\begin{figure}
   \centering
   \includegraphics[width=\columnwidth]{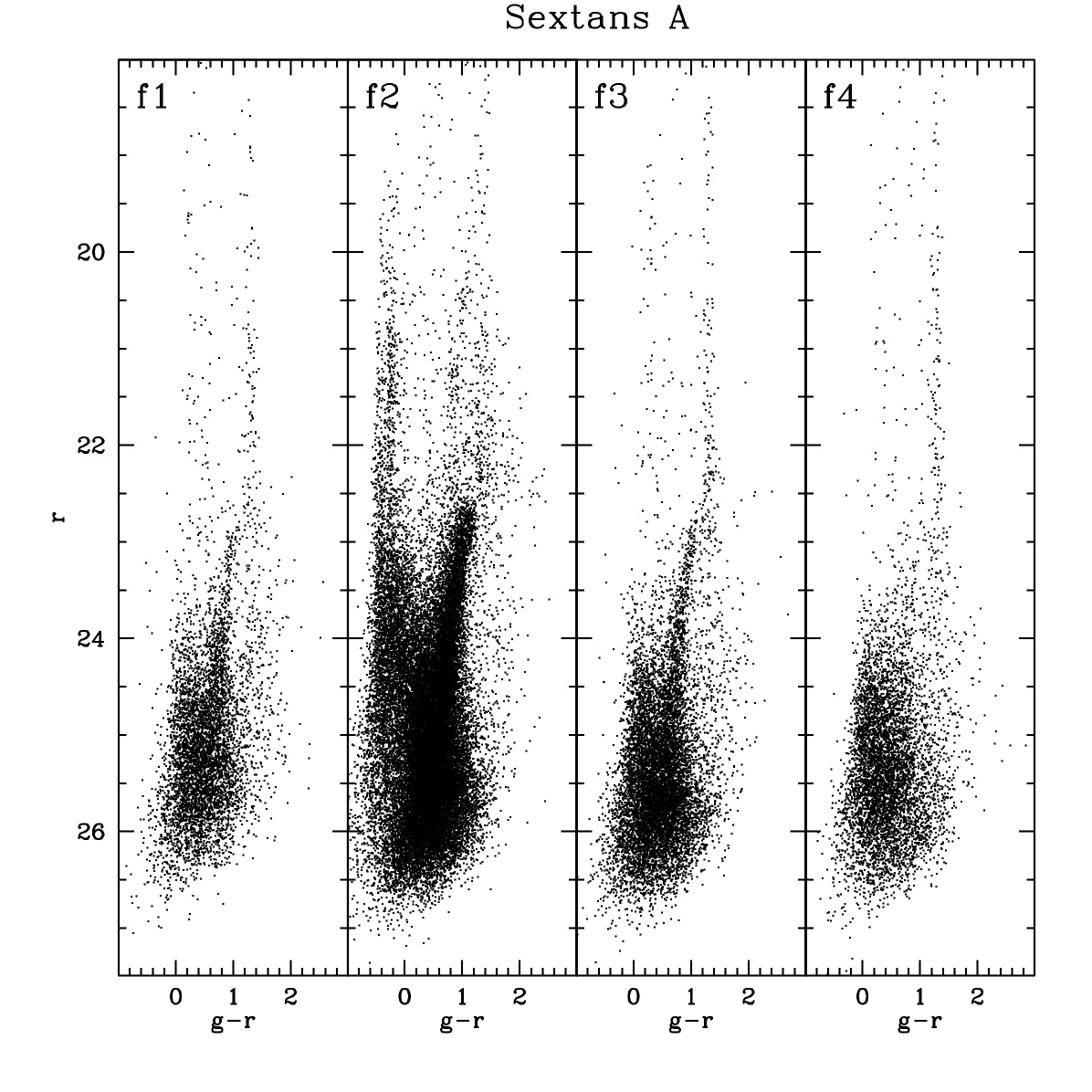}
   \includegraphics[width=\columnwidth]{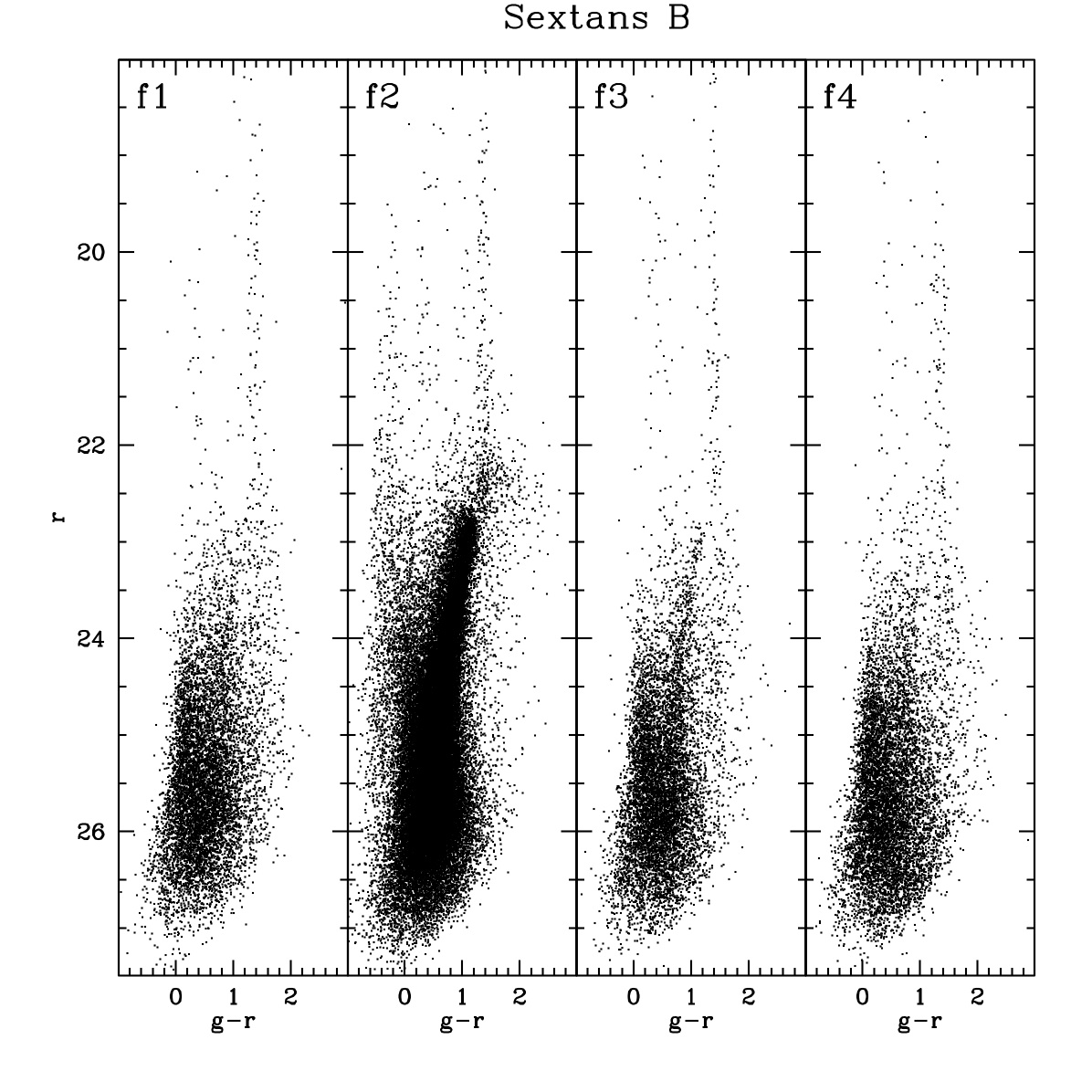}
     \caption{Chip-by-chip color magnitude diagrams of Sex~A (upper panels) and Sex~B (lower panels).}
        \label{cmds}
    \end{figure}

In Fig.~\ref{cmds} we present our final chip-by-chip Color Magnitude Diagrams (CMD) for Sex~A and Sex~B. The quality of the CMDs is similar to Pap-I; also in the present case we barely reach the Red Clump level in the outermost, less crowded regions. As known from previous studies \citep[in particular][]{dohm_cmd,dal09} the CMDs of the two galaxies are pretty similar: they are dominated by  a strong and wide Red Giant Branch (RGB), ranging from (r, g-r)$\sim(27.0,0.4)$ to $\sim(22.5,1)$. Above the RGB Tip a sprout of bright AGB stars (revealing the presence of an intermediate-age population) can be seen, more abundant in Sex~B than in Sex~A. On the other hand, the young Main Sequence (MS) plume at the blue edge of the CMDs is much more prominent in Sex~A, where the separation between H-burning MS and He-burning Blue Loop (BL) stars is also evident for $r\le 23.0$, MS stars populating the bluer of the two parallel features. A plume of Red Super Giants (RSG) is also evident around g-r$\sim 1.0$, from r$\sim 22.5$ to r$\sim 20.0$ \citep[see][for a detailed discussion and identification of all these features in the CMD of Sex~A; for a comparison with HST CMDs see Appendix~\ref{quality}.]{dohm_sfh}. 

At odds with the case presented in Pap-I, the stellar body of both galaxies exceeds f2: the characteristic RGB is clearly present also in f1 and f3, while f4 can be taken as a reasonable representation of the fore/background contamination affecting our CMDs. Following the detailed discussion presented in Pap-I it is easy to identify (especially in both the f4 CMDs) the thin vertical plume 
spanning the whole diagram around g-r$\simeq 1.5$, due to local M dwarfs, and the sparse wide band around g-r$\simeq 0.6$, bending to the red for $r\ga 22.0$, due to foreground MS stars in the Galactic halo, while the broad blob of sources
with $0.0\la$ g-r$\la 1.0$ and $r\ga 23.5$ is largely dominated by distant unresolved  galaxies (see Pap-I for a detailed discussion).

\begin{table}
  \begin{center}
  \caption{Observed and derived parameters of Sextans~A}
  \label{Tab_parA}
  \begin{tabular}{lcr}
    \hline
    Parameter & value & notes\\
\hline
$\alpha_0$     & 10:11:00.1                     & J2000$^a$  \\
$\delta_0$     & -04:41:42.6                    & J2000$^a$ \\
$l_0$          & $246.15\degr$                  & Gal. long.\\
$b_0$          & $39.87\degr$                   & Gal. lat.\\
SGL            & $109.00\degr$                   & Supergal. long.\\
SGB            & $-40.66\degr$                  & Supergal. lat.\\
E(B-V)         & $0.035\pm 0.002$               & Average $\pm \sigma^b$ \\
$(m-M)_0$      & $25.77\pm 0.12$                &   \\
D              & $1.42\pm 0.08$ Mpc             &   \\
1~arcsec       & 6.9~pc                         & conv. factor at D=1.42~Mpc\\
$V_{\rm h}$    & $324\pm 1$~km~s$^{-2}$         & heliocentric velocity$^c$ \\
$V_{g}$        & $163$~km~s$^{-1}$              & galactocentric velocity$^c$ \\
$\langle{\rm [Fe/H]}\rangle$& $ -1.45$       & from RGB color$^d$ \\
$\epsilon$     & $0.0$                         & adopted  \\
PA             & $0\degr$                      & adopted  \\
$\mu_V(0)$     & $23.9 \pm 0.1$ mag/arcsec$^2$ & central SB$^e$   \\ 
$h$          & $2.5\arcmin$                  & S\'ersic scale radius \\
$r_h$        & $1.8 \arcmin$                   & Observed$^f$  \\
$V_{tot}$      & $11.7 \pm 0.2$                &  Observed$^f$ \\
$M_V$          & $-14.2 \pm 0.3$               &   \\
$L_V$          & $4.1^{+1.3}_{-1.0}\times 10^7~L_{V,\sun}$    & total V luminosity \\
$M_{HI}$       & $6.2\times 10^7~M_{\sun}$      & gaseous mass$^g$  \\
$M_{dyn}$      & $\sim 1\times 10^9~M_{\sun}$      & total mass$^h$\\
\hline
\end{tabular}
\tablefoot{$^a$ Estimated by eye from our images. 
$^b$ Average over the whole FoV from the reddening maps by \citet{ebv}, as re-calibrated by 
\citet{schlaf}. 
$^c$ From \citet{mcc}.
$^d$ From synthetic CMD fit on HST photometry by \citet{dolph}. 
$^e$ Average of $\mu_V$ for $R_{\epsilon}<1.8\arcmin$. Not corrected for extinction. Note that $\mu_V(0)$ values extrapolated from fits with not-fully-adequate models (e.g. exponential models) can differ significantly from this {\em observed} value. 
$^f$ From the numerical integration of the SB profile in $r_{\epsilon}$.
$^g$ From \citet{ott}. 
$^h$ See Sect.~\ref{HI}.} 
\end{center}
\end{table}

\begin{table}
  \begin{center}
  \caption{Observed and derived parameters of Sextans~B}
  \label{Tab_parB}
  \begin{tabular}{lcr}
    \hline
    Parameter & value & notes\\
\hline
$\alpha_0$     & 10:00:00.05                    & J2000$^a$  \\
$\delta_0$     & +05:19:56.4                    & J2000$^a$ \\
$l_0$          & $233.20\degr$                  & Gal. long.\\
$b_0$          & $43.78\degr$                   & Gal. lat.\\
SGL            & $95.46\degr$                   & Supergal. long.\\
SGB            & $-39.62\degr$                  & Supergal. lat.\\
E(B-V)         & $0.024\pm 0.003$               & Average $\pm \sigma^b$ \\
$(m-M)_0$      & $25.80\pm 0.12$                &   \\
D              & $1.44\pm 0.08$ Mpc             &   \\
1~arcsec       & 7.0~pc                         & conv. factor at D=1.44~Mpc\\
$V_{\rm h}$    & $304\pm 1$~km~s$^{-1}$         & heliocentric velocity$^c$ \\
$V_{g}$        & $171$~km~s$^{-1}$              & galactocentric velocity$^c$ \\
$\langle{\rm [Fe/H]}\rangle$& -1.6              & from RGB color \\
$\epsilon$     & $0.31$                         & adopted ellipticity$^c$  \\
PA             & $95\degr \pm 15\degr$          & from $R\le150\arcsec$  \\
$\mu_V(0)$     & $23.1 \pm 0.1$ mag/arcsec$^2$ & central SB$^d$   \\ 
$h$          & $2.0\arcmin$                  & S\'ersic scale radius$^e$  \\
$r_h$        & $1.9 \arcmin$                   & Observed$^f$  \\
$V_{tot}$      & $11.7 \pm 0.1$                &  Observed$^f$ \\
$M_V$          & $-14.2 \pm 0.2$               &   \\
$L_V$          & $4.1^{+0.8}_{-0.7}\times 10^7~L_{V,\sun}$    & total V luminosity \\
$M_{HI}$       & $4.1\times 10^7~M_{\sun}$      & gaseous mass$^g$  \\
$M_{dyn}$      & $\sim 1\times 10^9~M_{\sun}$      & total mass$^h$\\
\hline
\end{tabular}
\tablefoot{$^a$ From \citet{cotton}. $^b$ Average over the whole FoV from the reddening maps by \citet{ebv}, as re-calibrated by \citet{schlaf}. $^c$ From \citet{mcc}.
$^d$ Average of $\mu_V$ for $r_{\epsilon}\le 0.5\arcmin$. Not corrected for extinction. Note that $\mu_V(0)$ values extrapolated from fits with not-fully-adequate models (e.g. exponential models) can differ significantly from this {\em observed} value.
$^e$ From the best-fit model for the entire profile (continuous line in Fig.~\ref{prof}).
$^f$ From the numerical integration of the SB profile in $r_{\epsilon}$.
 $^g$ From \citet{ott}.
 $^h$ See Sect.~\ref{HI}.} 
\end{center}
\end{table}

\subsection{Artificial stars experiments}
\label{comple}

The completeness of the stellar catalogs has been estimated by means of extensive artificial stars experiments, limited to Chip~2, since in both the cases this field samples the whole range of crowding encountered in the total LBC FoV. Following the same procedure adopted in Pap-I, a total of 
$\sim 100000$ artificial stars per galaxy have been added to the images  and the entire data reduction process has been repeated as in the real case, also adopting the same selection criteria described above. The PSF adopted as the best-fit model for photometry was also assumed as the model for the artificial stars. Artificial stars were distributed uniformly in position, over the entire extent of f2, and in color, over the range $-0.6\le g-r\le 2.0$. They were distributed in magnitude according to a luminosity function similar to the observed 
one but monotonically increasing also beyond the limit of the photometry, down to $r\simeq 27.5$ \citep[see][for details and discussion]{lf}.

   \begin{figure}
   \centering
   \includegraphics[width=\columnwidth]{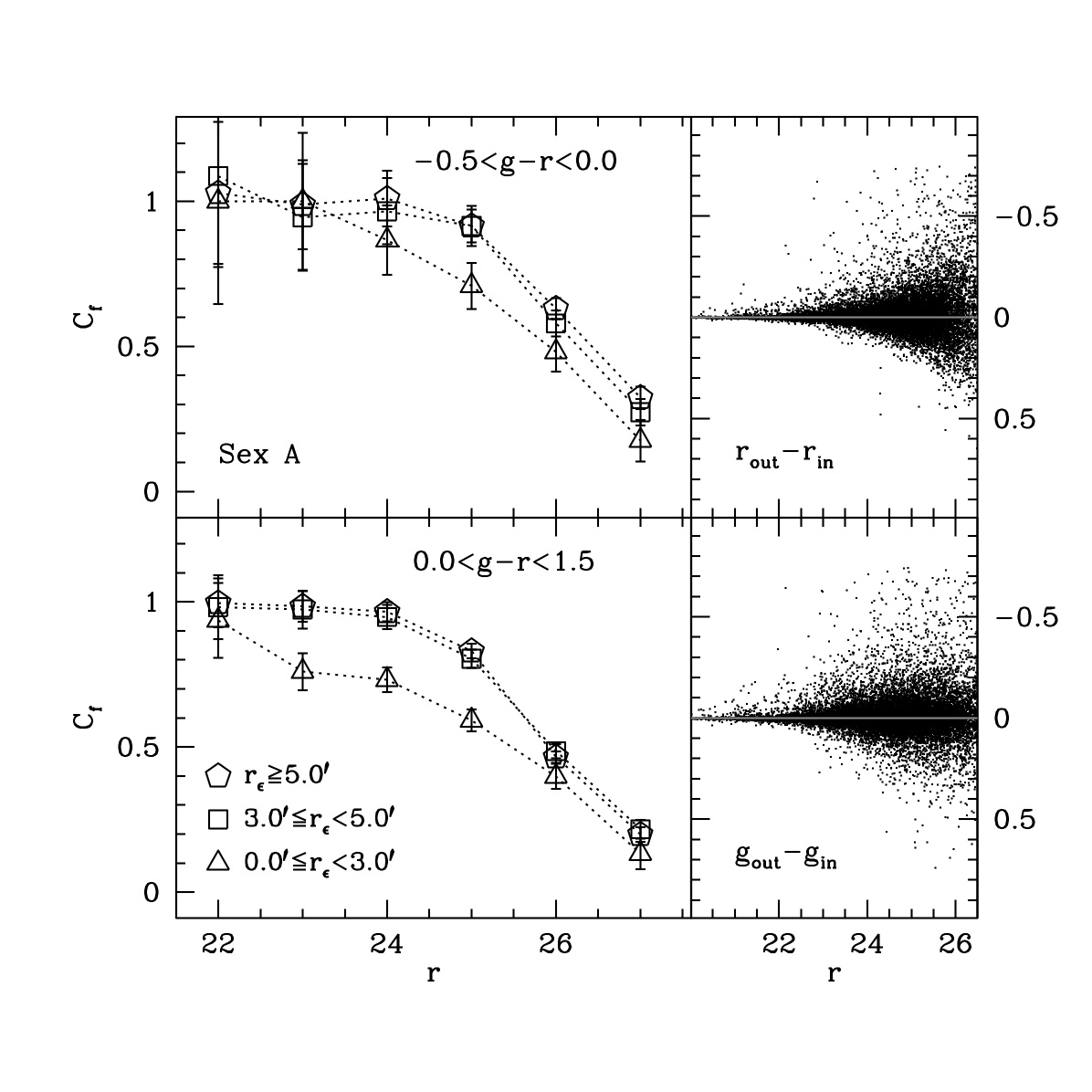}
   \includegraphics[width=\columnwidth]{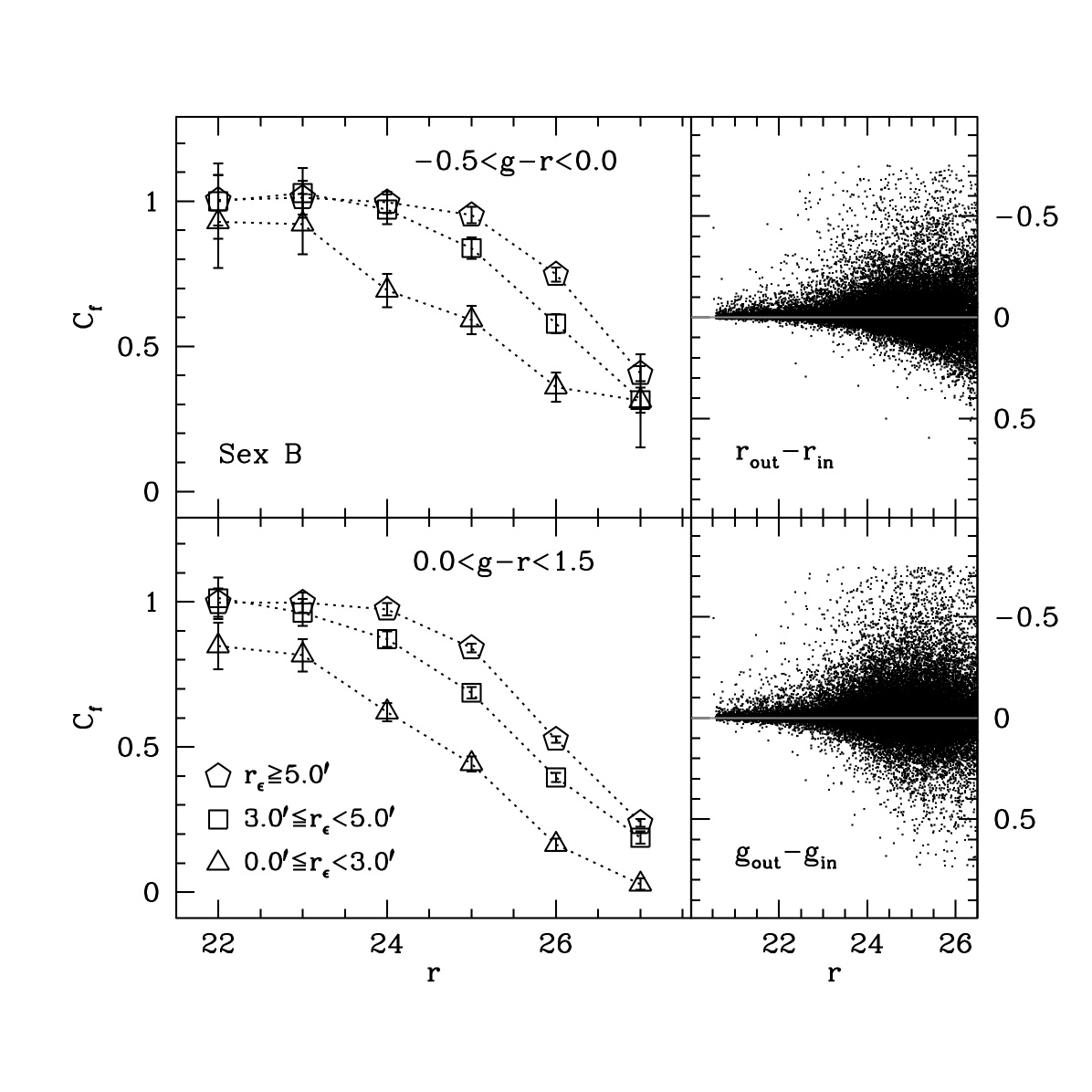}
     \caption{Completeness fraction as a function of r magnitude for different color ranges (upper and lower panels) and for different (elliptical) radial ranges (different symbols), for Sex~A (upper panel) and Sex~B (lower panel).}
        \label{CfA}
    \end{figure}


In Fig.~\ref{CfA}, we show the completeness fraction ($C_f$) as a function of r magnitude for different color ranges and for different (elliptical) radial ranges (see Sect.~\ref{sbp}, for a definition of $r_{\epsilon}$), and the
difference between input and output magnitudes  for all the artificial stars recovered. The significantly higher degree of crowding in the inner regions of Sex~B, with respect to Sex~A, is at the origin of the difference in the radial behavior of the completeness as well as in the (relative) photometric accuracy. As quantitative reference, in Table~\ref{Tab_acc} we 
provide the standard deviations of the input-output magnitude differences from artificial stars in various magnitude bins.
These are average values over the whole f2 fields, the actual accuracy depending on the distance from the central regions of the galaxies.

\begin{table}
  \begin{center}
  \caption{Accuracy of the relative photometry from artificial stars experiments$^a$.}
  \label{Tab_acc}
  \begin{tabular}{lcccc}
    \hline
                &  Sex~A     &            &  Sex~B     &           \\  
 mag range      & $\sigma_g$ & $\sigma_r$ & $\sigma_g$ & $\sigma_r$\\
$r<22.0$        & 0.02       &  0.01      & 0.03       & 0.02      \\
$22.0\le r<23.0$& 0.05       &  0.03      & 0.06       & 0.05      \\
$23.0\le r<24.0$& 0.06       &  0.04      & 0.08       & 0.06      \\
$24.0\le r<25.0$& 0.09       &  0.07      & 0.11       & 0.08      \\
$25.0\le r<26.0$& 0.12       &  0.12      & 0.15       & 0.12      \\
\hline
\end{tabular}
\tablefoot{$^a$ Averaged over the whole field f2 of each set of observations.} 
\end{center}
\end{table}

\subsection{Surface photometry and coordinate system}
\label{sbp}

As a basis for the construction of the final composite  profiles 
(surface photometry + star counts; see Sect.~\ref{struc}) we adopted the surface photometry by \citet[][HE06, hereafter]{HE06}, kindly provided by D. Hunter. HE06 surface photometry extends to relatively large radii (198$\arcsec$ and 280$\arcsec$, for Sex~A and Sex~B, respectively) and it is carefully corrected for the contribution of several bright foreground stars superposed to the main body of the two galaxies. 
For sanity check we also performed surface photometry in the innermost $144\arcsec$ of the best $g$ and $r$ images using XVISTA\footnote{\tt http://astronomy.nmsu.edu/holtz/xvista/index.html} 
\citep[see][and Pap-I, for details on the code and on the adopted procedure]{xvista,lucky}.
In the overlapping regions the shapes of the profiles obtained with XVISTA from our images and those by HE06 are in excellent agreement.  

Inspecting the ellipticity ($\epsilon$) and  position angle (PA) profiles
we decided (a) to consider Sex~A as round, as a basic approximation for the purpose of deriving the SB profile, given its low ellipticity and, consequently, poorly 
determined PA ($\epsilon=0.17$ and $PA=0\degr$, according to M12; $\epsilon=0.15$ and $PA=41.8\degr$, according to HE06), (b) to adopt, for Sex~B, $PA=95\degr\pm 15\degr$, from the average of the 
g and r profiles within $R\le 144\arcsec$ obtained with XVISTA, in reasonable agreement with the estimates reported by M12 ($PA=110\degr$) and HE06 ($PA=87.6\degr$), and, (c) for the same galaxy, to adopt $\epsilon=0.31$ from M12, as we found it more appropriate over the whole body of the galaxy than the typical value  we obtained with XVISTA, $\epsilon\simeq0.4$ and estimated by HE06 ($\epsilon\simeq0.41$). All these choices, as well as the position of the centers of the galaxies, have been verified to be adequate for the stellar components of the galaxies (and preferable to values found in the literature) by detailed visual inspection of our images, and comparison with superposed ellipses with the proper parameters. For the shape parameters of the \HI~ components see Sect.~\ref{HI}. It must be recalled that these galaxies have, by definition, somehow {\em irregular morphologies} and the adoption of a single value of 
$\epsilon$ and $PA$ over the whole extension of the galaxies is just a convenient approximation that allows us to parametrize the overall morphology with a simple model (see, e.g., HE06 for discussion).

Assuming the coordinates for the center listed in Tab.~\ref{Tab_parA} and Tab.~\ref{Tab_parB}, we
convert to projected cartesian coordinates $X$,$Y$(in arcmin) and adopt the elliptical radius
$r_{\epsilon}$, equivalent to the major-axis radius, as the radial coordinate of reference
(see Pap-I for references and definitions).

\section{Population gradients}
\label{grad}

\begin{figure}
   \centering
   \includegraphics[width=\columnwidth]{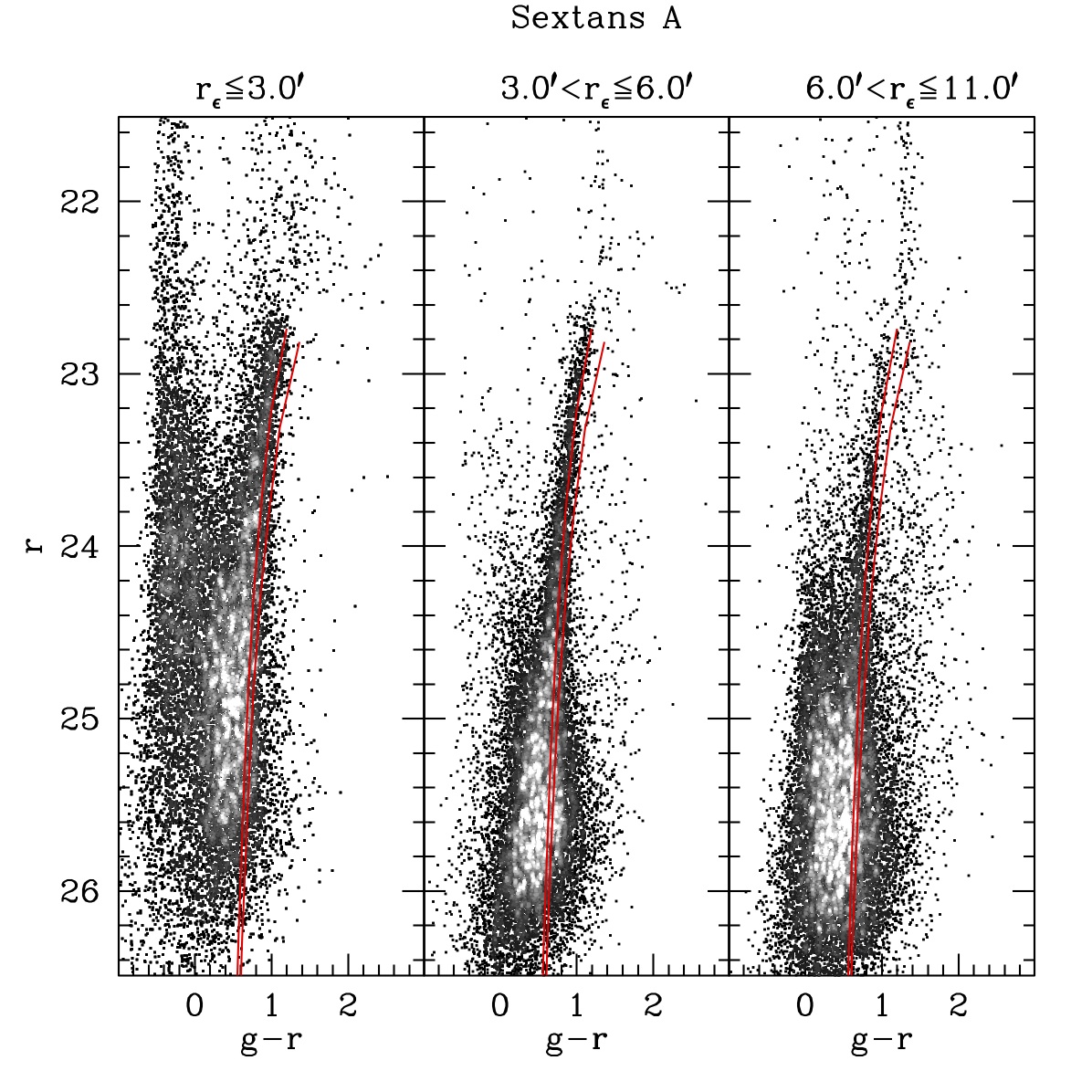}
   \includegraphics[width=\columnwidth]{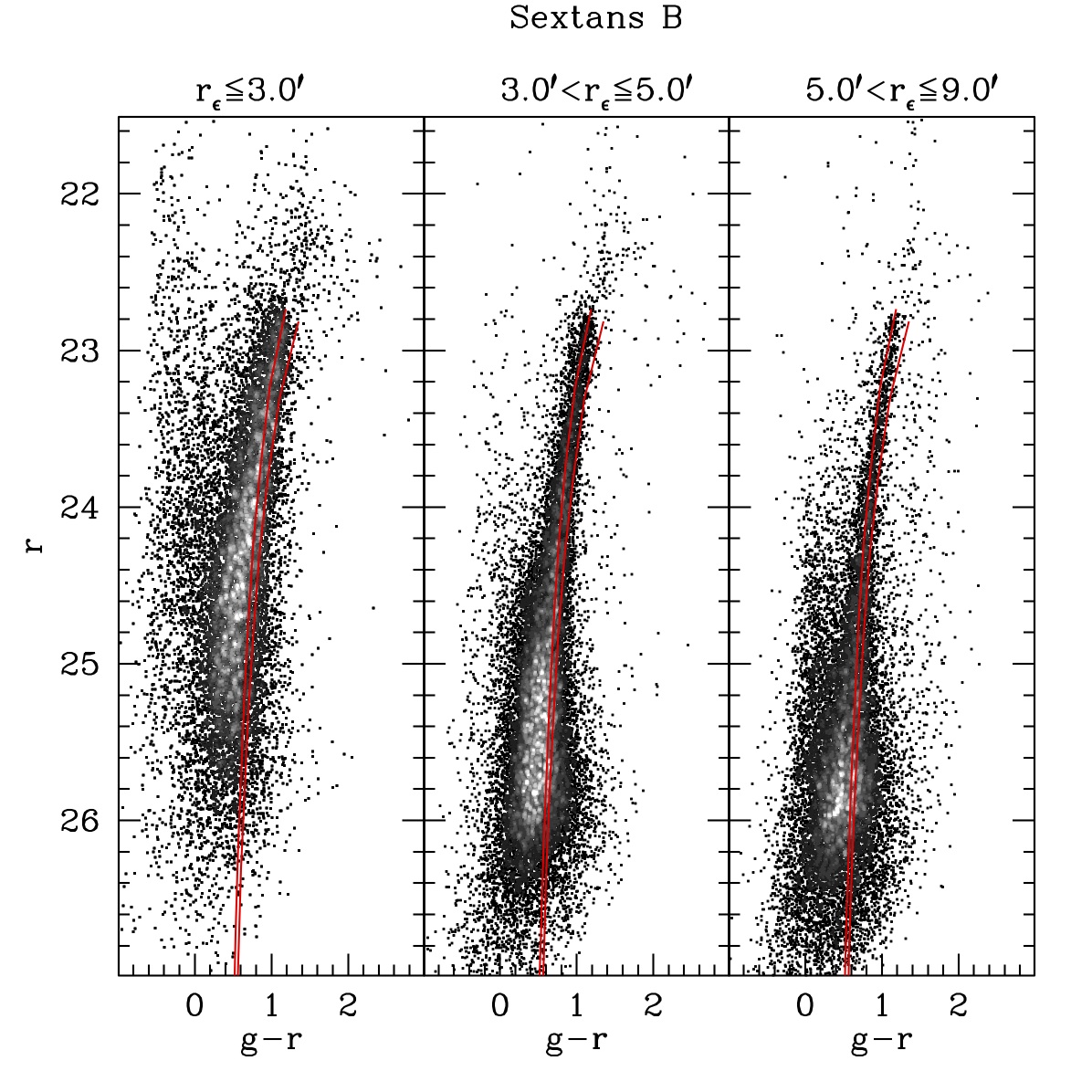}
     \caption{CMDs for Sex~A (upper panels) and Sex~B (lower panels) in different radial ranges. The red giant branches are compared with the ridge lines of the globular clusters M92 ([Fe/H]=-2.31) and M3 ([Fe/H]-1.50), from left to right. In all the panels stars are plotted as black points in regions of the CMD with a few stars and as gray squares otherwise, with the scale of gray proportional to the local density in the CMD. Lighter tones of gray correspond to higher density.}
        \label{cmrad}
    \end{figure}

In Fig.~\ref{cmrad} we compare the CMDs of Sex~A and Sex~B in different radial ranges with the ridge lines of two metal-poor globular clusters: M92 ([Fe/H]=-2.31) and M3 ([Fe/H]-1.50; see \citealt{harris,euge}).
The ridge lines are from the set by \citet{clem}, converted from g$\prime$,r$\prime$ to g,r according to \citet{tuck}, as in Pap-I. The assumed reddening and distance moduli of the clusters are from the 2010 version of the \citet{harris} catalog. 

The radial thresholds have been chosen (a) to clearly separate the innermost region - containing most of the young stars and being maximally affected by the crowding - from the rest of the galaxy, where the quality of the photometry and the completeness varies only very mildly with radius, and (b) to look for differences in the stellar populations in correspondence to breaks in the Surface Brightness (SB) profiles (see Sect.~\ref{struc}). The overall extent of the SB profiles drove also the choice of the outermost $r_{\epsilon}$ limit considered in Fig.~\ref{cmrad}.

The basic conclusion that can be drawn from Fig.~\ref{cmrad} is that, in both galaxies, there is no obvious difference in the color distribution of RGB stars in the different radial ranges considered. The old to intermediate-age population traced by these stars have colors typical of metal-poor RGB stars (in general [Fe/H]$\la -1.5$) at any radius in both galaxies. The RGB appears slightly bluer in Sex~A than in Sex~B, suggesting a lower mean metallicity or/and a lower mean age in the first galaxy. It must be recalled that the absolute calibration of our Sex~A photometry is more uncertain, and this may play a role in this context. Still, the RGB of Sex~A lies on the blue side of the RGB of Sex~B also when the HST CMDs are compared, strongly suggesting that  the difference observed in Fig.~\ref{cmrad} is real.
Indeed \citet{weisz} found that Sex~B formed $>50$\% of its stars more than 12.5~Gyr ago, while Sex~A reached the same fraction of star formation only 5~Gyr ago, pointing to a younger average age as the main driver of the difference in the RGB color between the two galaxies. 

Concerning the metallicity, M12 reports $\langle {\rm [Fe/H]}\rangle\sim -1.85$ for Sex~A, from the mean V-I color of the RGB determined from the shallow photometry by \citet{sakai}, while from the fit of their HST CMD \citet{dolph} estimated $\langle {\rm [M/H]}\rangle= -1.45$, with a spread of $\pm 0.20$ dex. 

On the other hand, no abundance estimate for the old to intermediate-age of Sex~B is available. Given the large fraction of very old stars that should populate the RGB of Sex~B \citep[according to the star formation history derived by][]{weisz}, the color of this sequence should provide a relatively reliable indication of the metallicity, when compared with old-age templates. Fig.~\ref{cmrad} implies that the vast majority of RGB stars in Sex~B should have $-2.3\la {\rm [Fe/H]}\la -1.5$. We obtain individual estimates for the RGB stars of Sex~B in the ANGST HST sample lying within one magnitude from the tip, by interpolating on a grid of globular clusters ridge lines as done in \citet{umi}. From this set of individual metallicities we obtain a mean metallicity $\langle{\rm [Fe/H]}\rangle=-1.6$, with a standard deviation of 0.35~dex, in the \citet{cg97} metallicity scale.

\citet{kniazev} and \citet{magrini} derived abundances estimates for the young populations in both galaxies from spectroscopic analysis of a few HII regions, finding typical metallicity values around 0.05 - 0.1 solar. \citet{kaufer} derived spectroscopic metallicity for three RSG in Sex~A, finding a mean [Fe/H]$\simeq -1.0$, in reasonable agreement with the above results, and [$\alpha$/Fe]$\simeq -0.1$.
The comparison between the typical metallicity of the oldest stars (RGB, [Fe/H]$\la -1.5$) and the youngest stars (OB stars in HII regions, Fe/H]$\sim -1.0$) implies a moderate degree of chemical enrichment over the lifetime of the galaxies.

   \begin{figure}
   \centering
   \includegraphics[width=\columnwidth]{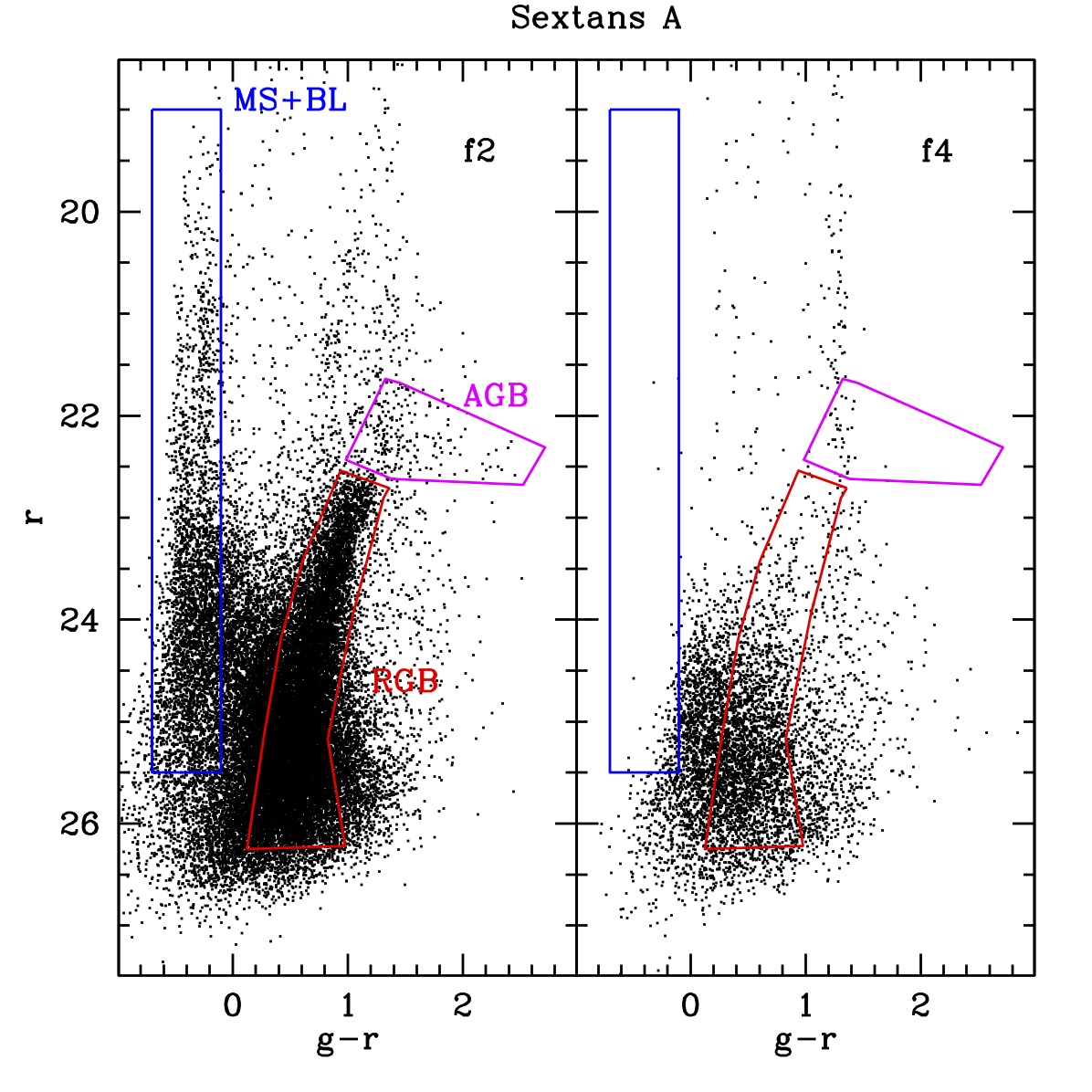}
     \caption{Boxes adopted to select stars in different evolutionary phases are superposed to
     the CMDs of a field dominated by Sex~A (f2) and a field dominated by the back/foreground population
     in its surroundings (f4).}
        \label{selbox}
    \end{figure}

\begin{figure}
   \centering
   \includegraphics[width=\columnwidth]{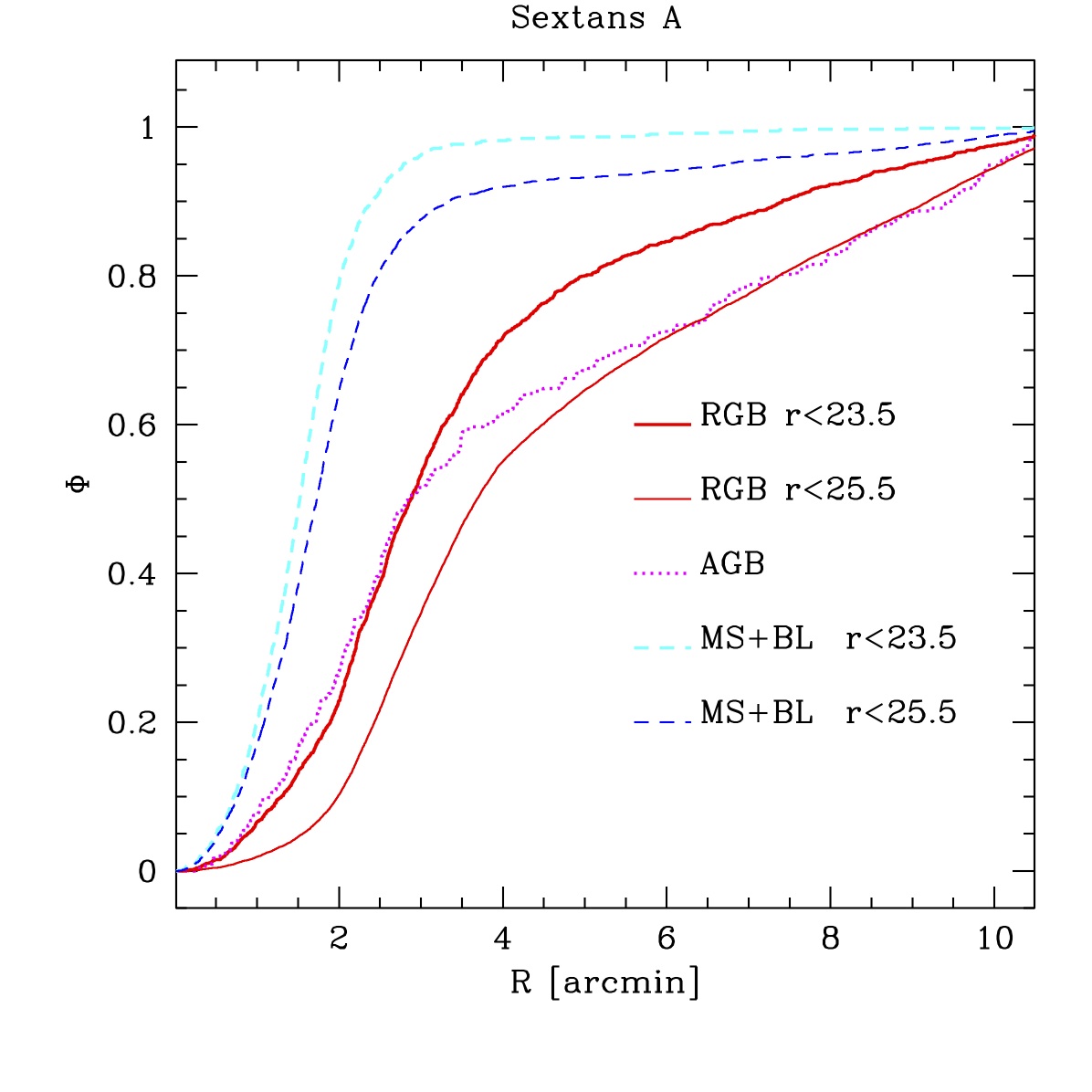}
   \includegraphics[width=\columnwidth]{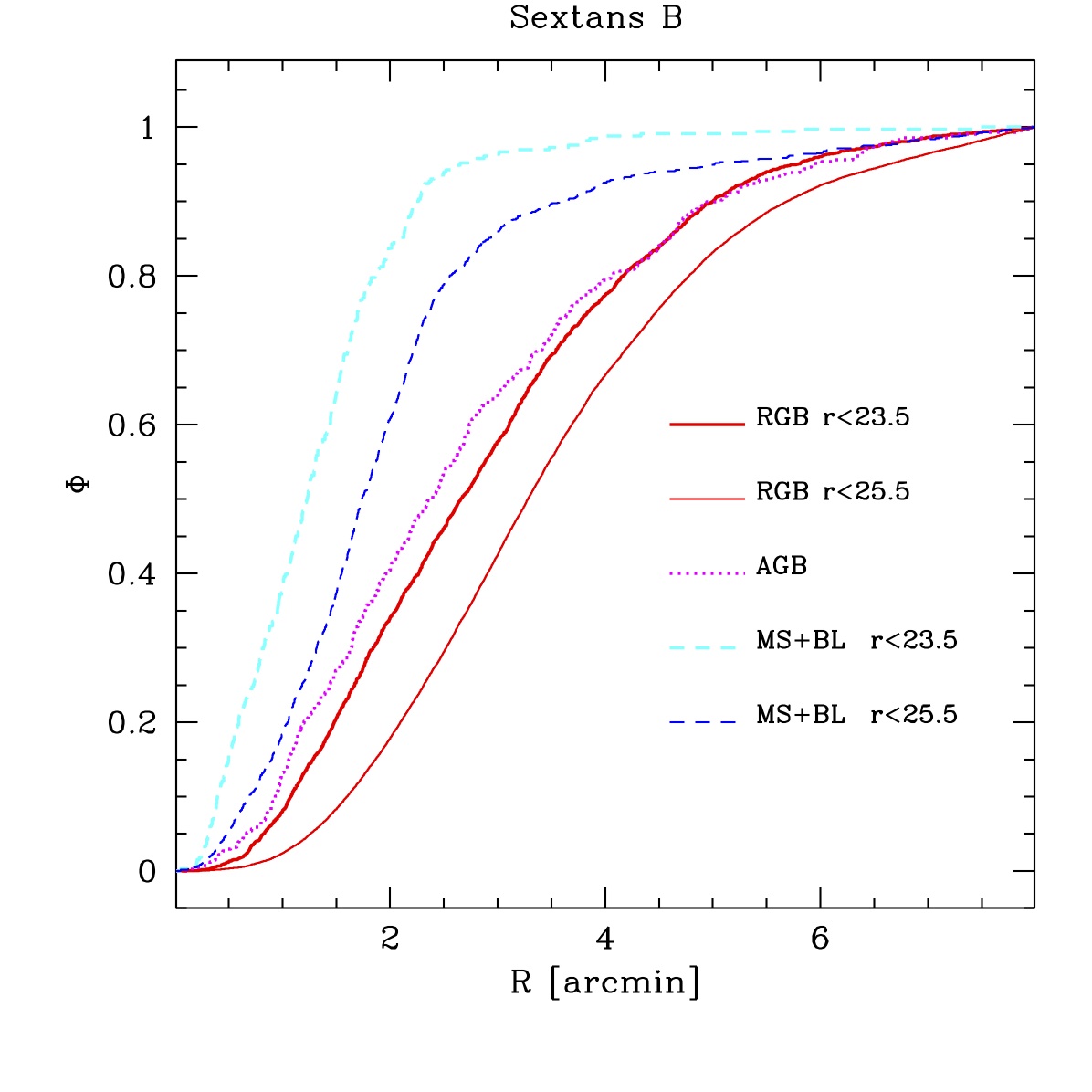}
     \caption{Cumulative radial distribution of different stellar species, selected according to Fig.~\ref{selbox}, for Sex~A (upper panel) and Sex~B (lower panel). Since this is a comparison between different populations within the same galaxy, we consider the distributions in the circular radius R, for simplicity.}
        \label{dist}
    \end{figure}

In the following of this Section we provide a first broad look at the relative spatial distribution of stars tracing different epochs of star formation. A more detailed analysis, including also the clustering properties of the young populations \citep[that are known to be fairly complex, at least in Sex~A, see][]{dohm_sfh}, also in relation with the distribution of HI, is deferred to a future contribution.
In Fig.~\ref{selbox} we show the boxes we adopt (in both galaxies) to select (a) MS and BL stars, tracing the youngest stellar populations (age$\la 300$~Myr), (b) bright AGB stars, tracing itermediate-age populations (up to age$\sim 5$~Gyr), and (c) RGB stars, tracing intermediate/old populations (age between $\sim 2$ and $\sim 13$ Gyr). In the right panel of the figure we superpose the same boxes to a CMD sampling the fore/background population, to provide a direct idea of the effect of contamination on the radial distribution of the various tracers. The MS+BL box is virtually unaffected by contamination. The effect should be negligible also for the RGB box, for r$<24.0$; on the other hand the sparse population of AGB stars can be significantly contaminated by foreground M dwarfs even at the center of the considered galaxies. Therefore, it must be taken into account that the radial distributions of AGB stars presented here are the convolution of the distribution of genuine AGBs and of local contaminants that are uniformly distributed (with the additional effects of holes and gaps). As a result the {\em true} radial distributions of AGB stars should be more centrally concentrated than what shown here.
All the comparisons displayed in Fig.~\ref{dist} are between sub-samples with the same cut at faint magnitudes, i.e. that have approximately the same degree of completeness; this condition is guaranteed for the $r<23.5$ samples (see Fig.~\ref{CfA}).

In Fig.~\ref{dist} it is apparent that in both galaxies the MS+BL population is much more concentrated than all the other stellar species considered in the comparison; in particular they are confined in the innermost $\simeq 3\arcmin \simeq 1.3$~kpc. The effect is clearer and more robustly determined for stars brighter than $r=23.5$, for the reasons mentioned above. 

AGB stars appear as concentrated as (Sex~A) or more concentrated than (Sex~B) bright RGB stars, in spite of the effects of contamination. Hence, both Sex~A and Sex~B appear to follow the general trend observed in all dwarf galaxies  \citep[see, e.g.][]{harbeck,tht}, i.e. younger/more metal rich populations are preferentially found near the center, while older (and more metal-poor) populations have extended distributions and dominate in the outermost regions.

\section{Structure}
\label{struc}

\begin{figure}
   \centering
   \includegraphics[width=\columnwidth]{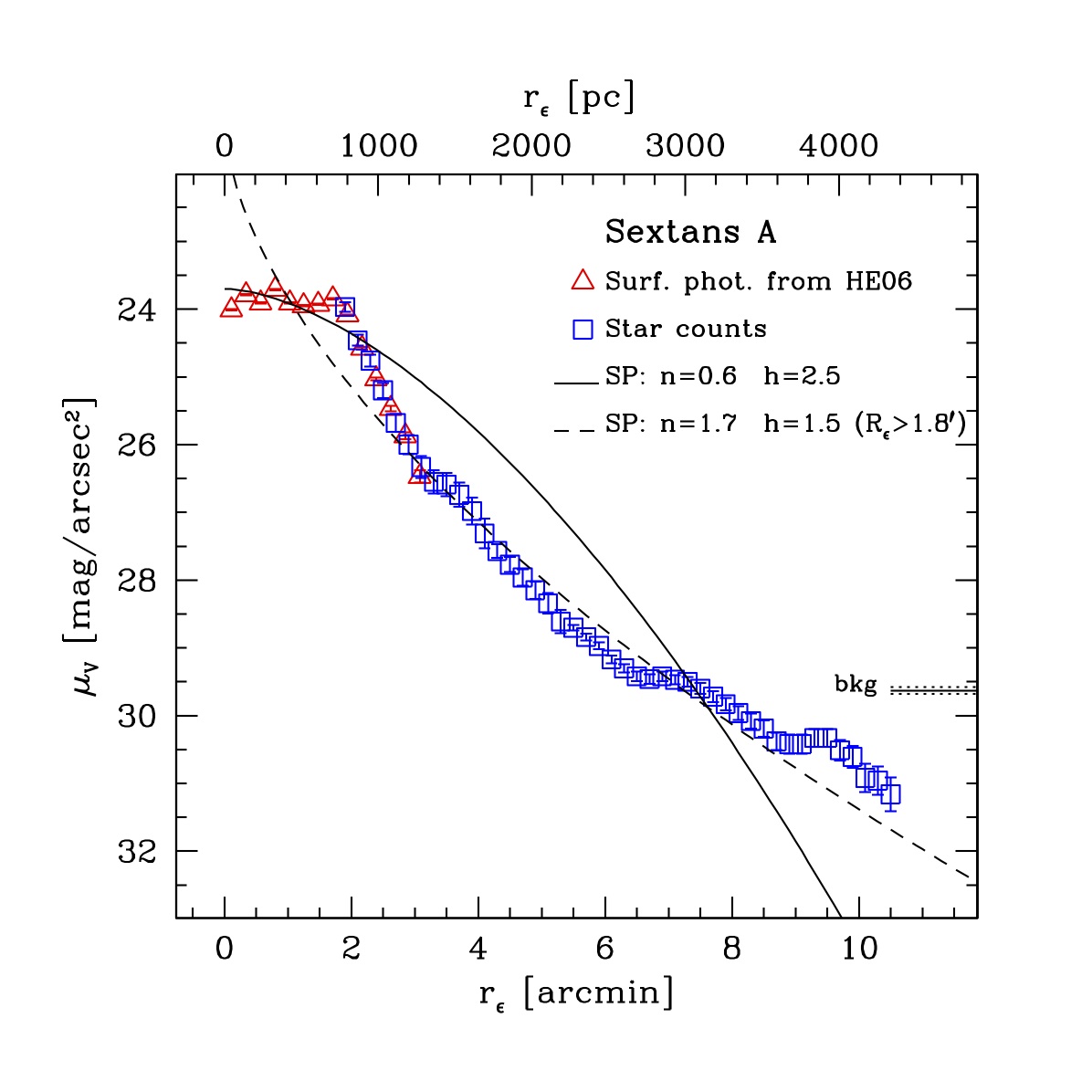}
   \includegraphics[width=\columnwidth]{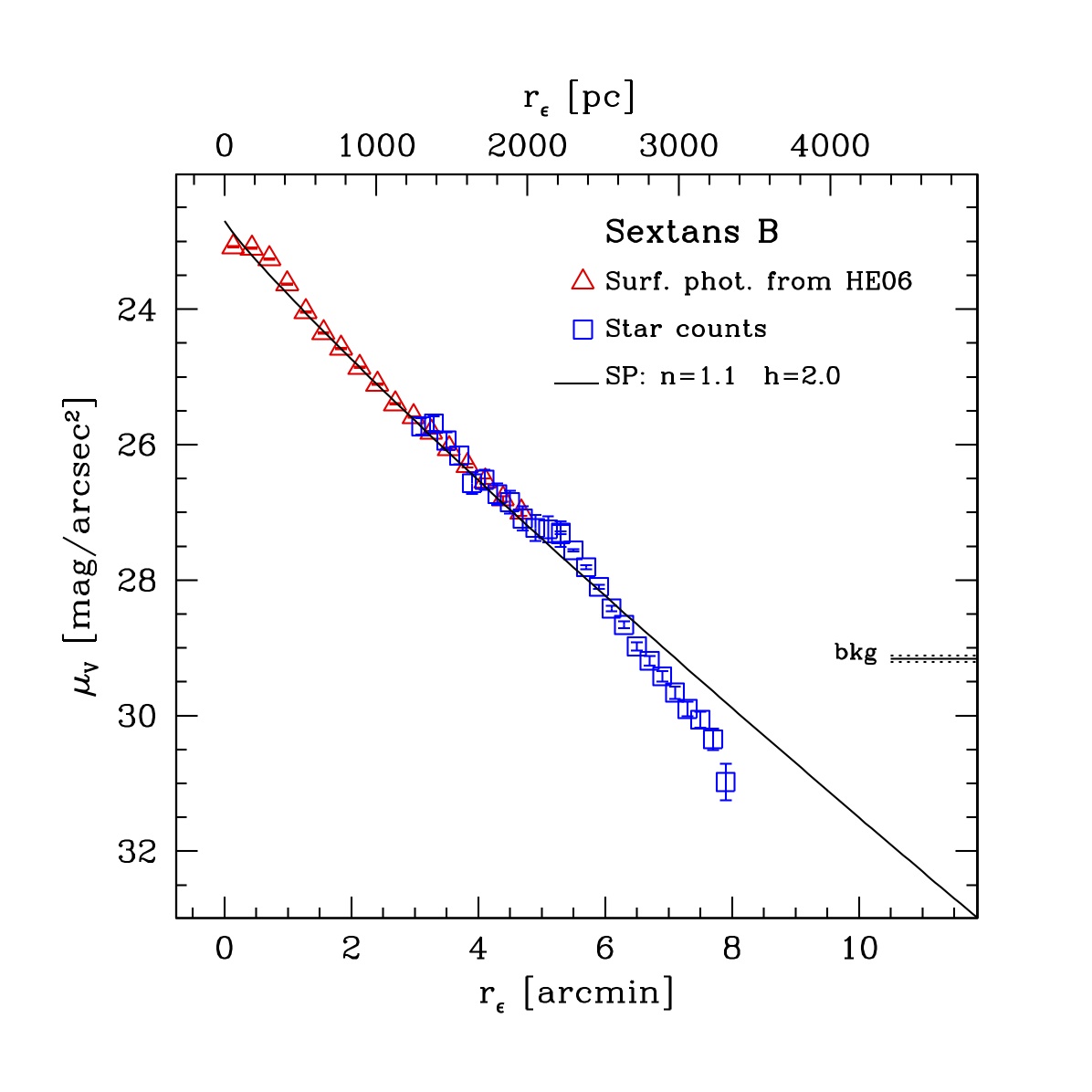}
     \caption{Surface brightness profiles of Sex~A (upper panel) and Sex~B (lower panel) in V band 
 obtained by joining surface photometry from HE06 (open triangles) and star counts (open squares) profiles. The continuous lines are the Sersic models that best-fits the profiles. The dashed line in the Sex~A plot is the Sersic model that best fits the observed profile for $R_{\epsilon}>1.8\arcmin$, i.e. excluding the flat portion of the profile near the center of the galaxy. The level of the background (bkg) and the associated uncertainty are also reported.}
        \label{prof}
    \end{figure}

In Fig.~\ref{prof} we show the azimuthally-averaged major-axis surface brightness (SB) profile of Sex~A and Sex~B. The profile has been obtained  by joining the V-band surface photometry from HE06  out to $r_{\epsilon}= 3.1\arcmin$ (for Sex~A) or $r_{\epsilon}= 4.7\arcmin$ (for Sex~B), with the surface density profile obtained from star counts  in elliptical annuli, keeping fixed the values of $\epsilon$ and $PA$ derived in Sect.~\ref{sbp}, \citep[see][and Pap-I, for a discussion of the procedure and details]{lucky}. 
The (large) overlap region between the profiles from surface photometry and from star counts  was used to normalize the star counts profile, shifting it to the same V magnitude scale of the surface photometry profile. 

For star counts we used, as homogeneous density tracers, candidate RGB stars enclosed in the polygonal box shown in Fig.~\ref{selbox}, except for the innermost portion of the Sex~A profile where we included also bright ($r\le 23.0$) MS and BL stars bluer than $g-r=0.0$. The background is estimated in wide areas in the f4 fields of the two samples, located as distant as possible from the centers of the galaxies.
The profiles are constructed by applying bright magnitude limits to the RGB selection in the inner regions and going to fainter limits in the outskirts, to get rid of variations of the completeness with distance from the galaxy center and to have the highest SB sensitivity in the low-SB external regions. The process starts by finding the upper limit in magnitude that selects a sample of tracer stars providing a profile with a large overlapping region with the surface photometry profile, where the slope of the two profiles is indistinguishable (e.g., the regions where triangle and squares overlaps in Fig.~\ref{prof}). This (a) guarantees that, from the beginning of the "same-slope" region, there is no radial variation in completeness for the considered sample (hence, the star-counts profile is reliable), and (b) allows a robust normalization of the star-counts profile to the photometric scale of the surface photometry profile. Then star-counts profiles from samples with fainter magnitude limits can be joined to the composite profile with the same criteria and techniques until the whole sample (RGB stars with $r\le 26.0$, in the present case) is considered. 
In both galaxies, the incompleteness (and its radial variation) in the innermost regions ($r_{\epsilon}\le 2\arcmin-3\arcmin$) is too severe (see, e.g., Fig.~\ref{CfA}) to trace the SB with star-counts; these region must be covered with the SB profile from surface photometry that, by definition, is not affected by incompleteness.

As in the case of VV124 presented in Pap-I, our newly derived composite profiles reaches much larger radii (by a factor of $\ga 2$) than previous studies, that were based solely on surface photometry.
In particular we are able to trace the stellar profiles out to $r_{\epsilon}\simeq 10.5\arcmin$ and $r_{\epsilon}\simeq 8.0\arcmin$ for Sex~A and Sex~B, respectively, showing that also these galaxies are much more extended than what previously observed. At present, all the isolated dwarfs for which deep wide-field photometry has been acquired with 8m class telescopes have been found to have SB profiles extending down to SB levels ($\ga 30$ mag/arcsec$^2$) that cannot be reached with ordinary surface photometry, at least for these nearby (resolved) galaxies \citep[see][Pap-I and this study]{vanse,sanna}.

\begin{table}
  \begin{center}
  \caption{Observed Surface Brightness profiles.}
  \label{sbprof}
  \begin{tabular}{lcclcc}
               & Sex A &           &              &   Sex B &      \\
$r_{\epsilon}$ & $\mu_V$ & source &$r_{\epsilon}$ & $\mu_V$ & source\\
arcmin         & mag/arcsec$^2$ & & arcmin & mag/arcsec$^2$ &\\
 0.11 & 24.01 $\pm$ 0.01 & HE06 & 0.14  & 23.08 $\pm$  0.01 &  HE06 \\
 0.34 & 23.79 $\pm$ 0.01 & HE06 & 0.43  & 23.10 $\pm$  0.01 &  HE06 \\
 0.57 & 23.91 $\pm$ 0.01 & HE06 & 0.71  & 23.26 $\pm$  0.01 &  HE06 \\
 0.80 & 23.71 $\pm$ 0.01 & HE06 & 0.99  & 23.63 $\pm$  0.01 &  HE06 \\
 1.03 & 23.91 $\pm$ 0.01 & HE06 & 1.28  & 24.04 $\pm$  0.01 &  HE06 \\
 1.25 & 23.95 $\pm$ 0.01 & HE06 & 1.56  & 24.35 $\pm$  0.01 &  HE06 \\
 1.48 & 23.93 $\pm$ 0.01 & HE06 & 1.84  & 24.58 $\pm$  0.01 &  HE06 \\
 1.71 & 23.85 $\pm$ 0.01 & HE06 & 2.13  & 24.86 $\pm$  0.01 &  HE06 \\
 1.90 & 23.97 $\pm$ 0.07 &  sc  & 2.41  & 25.11 $\pm$  0.01 &  HE06 \\
 2.10 & 24.46 $\pm$ 0.08 &  sc  & 2.69  & 25.40 $\pm$  0.01 &  HE06 \\
 2.30 & 24.76 $\pm$ 0.09 &  sc  & 2.98  & 25.59 $\pm$  0.01 &  HE06 \\
 2.50 & 25.20 $\pm$ 0.11 &  sc  & 3.10  & 25.73 $\pm$  0.12 &   sc  \\
 2.70 & 25.68 $\pm$ 0.13 &  sc  & 3.30  & 25.69 $\pm$  0.11 &   sc  \\
 2.90 & 26.00 $\pm$ 0.14 &  sc  & 3.50  & 25.94 $\pm$  0.12 &   sc  \\
 3.10 & 26.33 $\pm$ 0.16 &  sc  & 3.70  & 26.16 $\pm$  0.13 &   sc  \\
 3.30 & 26.55 $\pm$ 0.17 &  sc  & 3.90  & 26.57 $\pm$  0.16 &   sc  \\
 3.50 & 26.59 $\pm$ 0.17 &  sc  & 4.10  & 26.52 $\pm$  0.15 &   sc  \\
 3.70 & 26.74 $\pm$ 0.18 &  sc  & 4.30  & 26.73 $\pm$  0.16 &   sc  \\
 3.90 & 26.98 $\pm$ 0.20 &  sc  & 4.50  & 26.85 $\pm$  0.17 &   sc  \\
 4.10 & 27.31 $\pm$ 0.22 &  sc  & 4.70  & 27.09 $\pm$  0.18 &   sc  \\
 4.30 & 27.57 $\pm$ 0.10 &  sc  & 4.90  & 27.23 $\pm$  0.19 &   sc  \\
 4.50 & 27.77 $\pm$ 0.11 &  sc  & 5.10  & 27.25 $\pm$  0.19 &   sc  \\
 4.70 & 27.96 $\pm$ 0.12 &  sc  & 5.30  & 27.32 $\pm$  0.19 &   sc  \\
 4.90 & 28.15 $\pm$ 0.13 &  sc  & 5.30  & 27.30 $\pm$  0.02 &   sc  \\
 5.10 & 28.34 $\pm$ 0.15 &  sc  & 5.50  & 27.56 $\pm$  0.02 &   sc  \\
 5.30 & 28.61 $\pm$ 0.17 &  sc  & 5.70  & 27.81 $\pm$  0.03 &   sc  \\
 5.50 & 28.70 $\pm$ 0.04 &  sc  & 5.90  & 28.10 $\pm$  0.03 &   sc  \\
 5.70 & 28.84 $\pm$ 0.05 &  sc  & 6.10  & 28.42 $\pm$  0.04 &   sc  \\
 5.90 & 28.97 $\pm$ 0.05 &  sc  & 6.30  & 28.66 $\pm$  0.05 &   sc  \\
 6.10 & 29.17 $\pm$ 0.06 &  sc  & 6.50  & 28.98 $\pm$  0.06 &   sc  \\
 6.30 & 29.30 $\pm$ 0.06 &  sc  & 6.70  & 29.19 $\pm$  0.07 &   sc  \\
 6.50 & 29.42 $\pm$ 0.07 &  sc  & 6.90  & 29.42 $\pm$  0.08 &   sc  \\
 6.70 & 29.46 $\pm$ 0.07 &  sc  & 7.10  & 29.66 $\pm$  0.09 &   sc  \\
 6.90 & 29.42 $\pm$ 0.07 &  sc  & 7.30  & 29.90 $\pm$  0.11 &   sc  \\
 7.10 & 29.47 $\pm$ 0.07 &  sc  & 7.50  & 30.06 $\pm$  0.12 &   sc  \\
 7.30 & 29.50 $\pm$ 0.07 &  sc  & 7.70  & 30.35 $\pm$  0.16 &   sc  \\
 7.50 & 29.60 $\pm$ 0.08 &  sc  & 7.90  & 30.98 $\pm$  0.27 &   sc  \\
 7.70 & 29.72 $\pm$ 0.08 &  sc  &       &    		    &       \\
 7.90 & 29.83 $\pm$ 0.09 &  sc  &       &    		    &       \\
 8.10 & 29.96 $\pm$ 0.10 &  sc  &       &    		    &       \\
 8.30 & 30.08 $\pm$ 0.11 &  sc  &       &    		    &       \\
 8.50 & 30.19 $\pm$ 0.12 &  sc  &       &    		    &       \\
 8.70 & 30.38 $\pm$ 0.13 &  sc  &       &    		    &       \\
 8.90 & 30.42 $\pm$ 0.14 &  sc  &       &    		    &       \\
 9.10 & 30.42 $\pm$ 0.14 &  sc  &       &    		    &       \\
 9.30 & 30.33 $\pm$ 0.13 &  sc  &       &    		    &       \\
 9.50 & 30.33 $\pm$ 0.13 &  sc  &       &    		    &       \\
 9.70 & 30.51 $\pm$ 0.15 &  sc  &       &    		    &       \\
 9.90 & 30.61 $\pm$ 0.16 &  sc  &       &    		    &       \\
10.10 & 30.92 $\pm$ 0.21 &  sc  &       &    		    &       \\
10.30 & 30.96 $\pm$ 0.21 &  sc  &       &    		    &       \\
10.50 & 31.16 $\pm$ 0.25 &  sc  &       &    		    &       \\
\hline
\end{tabular} 
\tablefoot{HE06 = surface photometry from HE06; sc = star counts from the 
present analysis. The uncertainties from in the HE06 photometry has been 
approximated at the 0.01 mag level.}
\end{center}
\end{table}

The profiles of both Sex~A and Sex~B display discontinuities and cannot be satisfactorily fitted with a single \citet{sersic} profile. Sex~A shows a flat branch in the innermost $\simeq 1.8\arcmin$, that is consistently traced in our own XVISTA  photometry and by HE06, who classify this galaxy among the "Flat inner profile" dwarfs (see HE06 for discussion). We will briefly return to this flattening in the stellar SB in Sect.~\ref{HI}, since it appears to have a counterpart in the \HI~ profile. 

Beyond $r_{\epsilon}\simeq 1.8\arcmin$ the profile displays two additional changes of slope, a shallow (and possibly not significant) one in correspondence of a weak kink at $r_{\epsilon}\simeq 3\arcmin$, and a more pronounced flattening at $r_{\epsilon}\simeq 6\arcmin$. 
The outer, more significant, change of slope is associated with a probable tidal tail, described in Sect.~\ref{dmap}, below.
The observed profile has some similarity with that of another isolated dIrr that has been traced down to very low SB, Leo~A \citep{vanse}. 
We note that a single Sersic model\footnote{Here we adopt the formalism by \citet{ciotti}, with the SB profile given by $I(r)=I_0e^{-b\eta^{\frac{1}{n}}}$ (their Eq.~1), with $\eta=\frac{r}{h}$, $h$ being the half-light radius. The value of the constant $b$ is from their Eq.~25, that works well for $n\ga 1.0$.} is significantly inadequate to represent the observed profile, even if the flat inner region is excluded from the fit. 

The profile of Sex~B displays a single change of slope, a weak steepening at $r_{\epsilon}\simeq 5\arcmin$.
Breaks in the SB profile are common in the outer region of disc galaxies \citep[][HE06]{kruit}. According to \citet{pohlen}, the downbending profile, with steeper outer region (their Type~II), as we observe in Sex~B, is very common in late type spiral galaxies \citep[see][for a local example]{ann_m33}; the incidence of the class declines for earlier types. The position ($\sim 2.5$ scalelength) of the profile break observed in Sex~B is slightly more external than the typical values for for Type~II disc galaxies \citep[$\sim 2.0$ scalelength][HE06]{pohlen}.

Since these galaxies should have the broad structure of fluffy thick discs, it is clear that the inclination of their principal plane with respect to the line of sight should play a role in determining the shape of their SB profile as seen from the Earth. Unfortunately the inclination of both galaxies is poorly constrained by \HI~ data (see Sect.~\ref{HI}). Moreover it is not at all guaranteed that the main stellar body of the galaxies (especially the backbone of old stars) have the same morphology and/or inclination of the \HI~ disc.

Looking at Fig.~\ref{cmrad}, above, it can be appreciated that no obvious change in the stellar content occurs in correspondence of the profile breaks, neither in Sex~A or Sex~B, as anticipated in Sect.~\ref{grad}. 

Integrating numerically the profiles shown in Fig.~\ref{prof} we obtain integrated V magnitudes
$V_{tot}=11.7\pm 0.3$ and $V_{tot}=11.7\pm 0.2$, and half-light radii $r_h=1.8\arcmin$ and $r_h=1.9\arcmin$, for Sex~A and Sex~B, respectively, in reasonable agreement with the values reported in the literature (see, e.g., HE06, M12), that are based on much less extended profiles and/or fits with (single) exponential profiles.

   \begin{figure}
   \centering
   \includegraphics[width=\columnwidth]{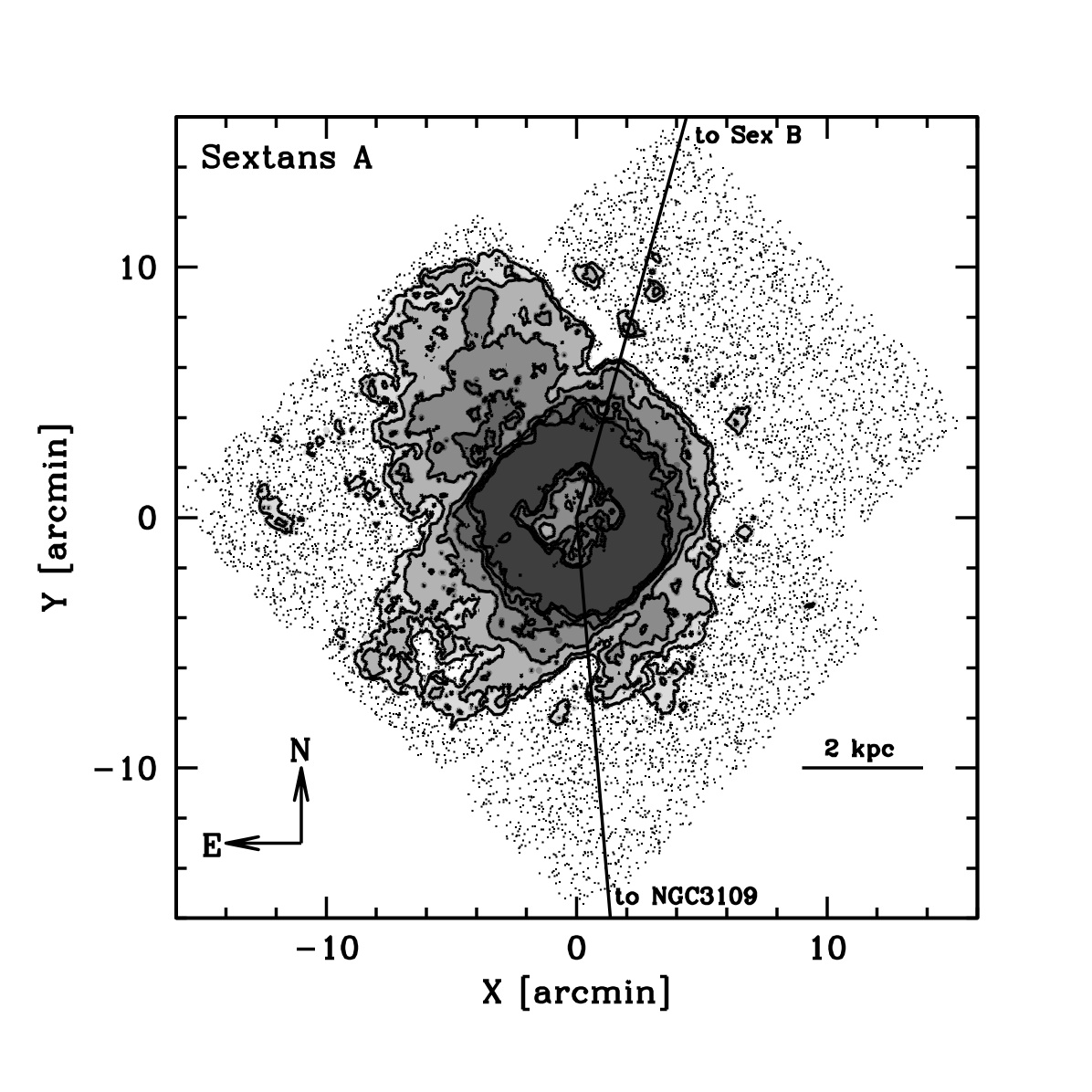}
   \includegraphics[width=\columnwidth]{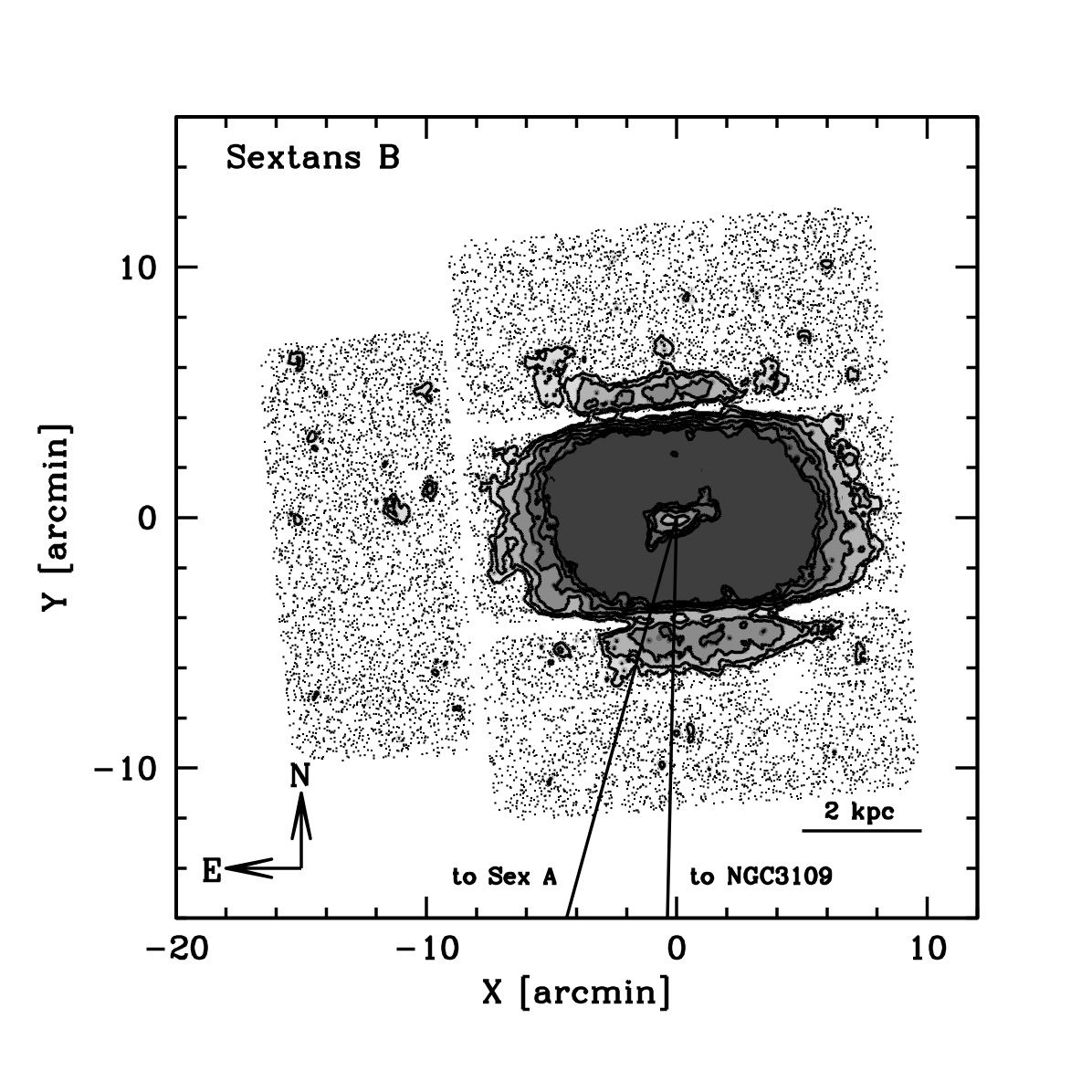}
     \caption{Stellar surface density contour maps from RGB counts. The contours / levels of grey correspond to density of 3, 5, 10, 20, and 40$\sigma$ above the background, from the lightest to the darkest tone of grey. 
     The density depressions near the center of the main galactic bodies are due to the low degree of completeness in the very crowded innermost regions.
The directions to nearby galaxies of the NGC~3109 group are plotted as continuous lines.
The distance to NGC~3109 is $\sim 500$~kpc and $\sim 700$~kpc, for Sex~A and Sex~B, respectively;
the distance between Sex~A and Sex~B is $\sim 250$~kpc.
All RGB stars are also plotted (as dots) to provide a direct illustration of the effects of the footprint of the camera and/or heavily saturated foreground stars. 
}
        \label{maps}
    \end{figure}


\subsection{Density maps}
\label{dmap}

In Fig.~\ref{maps} we present the density maps for Sex~A and Sex~B obtained from RGB star counts
on a fixed regular grid of nodes spaced by $0.1\arcmin$, with the same adaptive algorithm described in Pap-I\footnote{In Fig.~\ref{HImaps} we show the stellar density maps obtained with the matched filter technique \citep[see][]{connie} that was also used in Pap-I. As done in Pap-I, we present the maps obtained with both methods, as their full agreement  lends further support to the reliability of the discussed features.}. 
The lightest tone of grey corresponds to an over-density at $3\sigma$ above the background level (estimated in f4, as described in Sect.~\ref{struc}), and it is the outermost density contour that we can reliably trace with our data. Single isolated weak peaks with typical scales $\la 1\arcmin$ can correspond to clusters of galaxies whose red sequence galaxies fall in the RGB selection box. The small blob in the lower right corner of the northern chip in the Sex~B map is a good example that can be easily identified in the original images. However these features are very small compared to the contours of the galaxies and never exceed  $5\sigma$ above the background level, hence they cannot have a significant impact on the morphology of Sex~A and Sex~B as displayed in Fig.~\ref{maps}.
All the RGB stars selected for the analysis are also plotted in the maps, to give an idea of the sampling bias that are obviously affecting them (inter-chip gaps, saturated stars etc.). Also in this case the overall view of the main galactic bodies we are interested in is not compromised by these small-scale effects (albeit the discontinuities due to the gaps are clearly evident). Finally, the apparent density decline in the central region of the galaxies is due to the higher degree of crowding, and consequently lower completeness  (see Sect~\ref{comple}); in the case of Sex~A also the bright foreground star TYC~4907-713-1\footnote{A $V=11.5$ star located just $77.5\arcsec$ to the NE of the Sex~A center, that is heavily saturated in our images  (see Fig.~\ref{imaC}).} plays a role in preventing the detection of any  star over a region of several arcsec.
Again this is not a serious problem in the present context since the density in the central part is well described by the azimuthally averaged profiles, while here we are especially interested to the morphology of the most external regions of our galaxies.


The map of Sex~A reveals that the denser roundish main body is surrounded by a vast low-density and elongated envelope, where the density declines very gently. The feature is clearly visible also in the $10\sigma$ contours, hence there is no doubt that it is real. It is very interesting to note that the transition between the central body and the envelope occurs in correspondence to the outermost change of slope in the profile, at $R\simeq 6\arcmin$. The broad {\tt S}-shape of the outer envelope and its correspondence with a break toward a shallower slope in the density profile are suggestive of a tidal origin \citep{katy,pena,munoz}. In both maps of Fig.~\ref{maps} we report the direction to the nearest dwarf (Sex~B, in this case) and to NGC~3109, the most massive member of the association Sex~A and Sex~B belong to. No obviously significant alignment is apparent, but it must be reminded that the orientation of tidal tails depends significantly on the orbital phase \citep{klim_tid}.

At odds with its twin, Sex~B displays a remarkably elliptical but compact and regular morphology even in its outermost regions. Since the masses of the two galaxies are similar (see Sect.~\ref{HI}) the different morphology can be accounted for only by different initial conditions or different orbital histories, i.e. if Sex~A was more strongly affected than Sex~B by a past tidal interaction (see Sec.~\ref{disc}).

We will discuss in more detail the hypothesis of tidal interactions in the NGC~3109 group, also in relation with the results by \citet{penny} and \citet{filam} in Sect.~\ref{disc}.

Here, as in Pap-I, we use the following simple equation for the tidal radius \citep{keen_a,keen_b}

\begin{equation}
\label{RT}
r_t=\frac{2}{3}\left(\frac{m}{2M}\right)^{\frac{1}{3}} D_{1,2}
\end{equation}

which is appropriate for a logarithmic potential 
\citep[see][for discussion and references]{tidal}, to obtain some basic constraint on the encounter that 
can be at the origin of the putative tidal tail of Sex~A. $m$ and $M$ are the mass of the involved bodies (with $M>m$) and $D_{1,2}$ is their mutual distance.

   \begin{figure}
   \centering
   \includegraphics[width=\columnwidth]{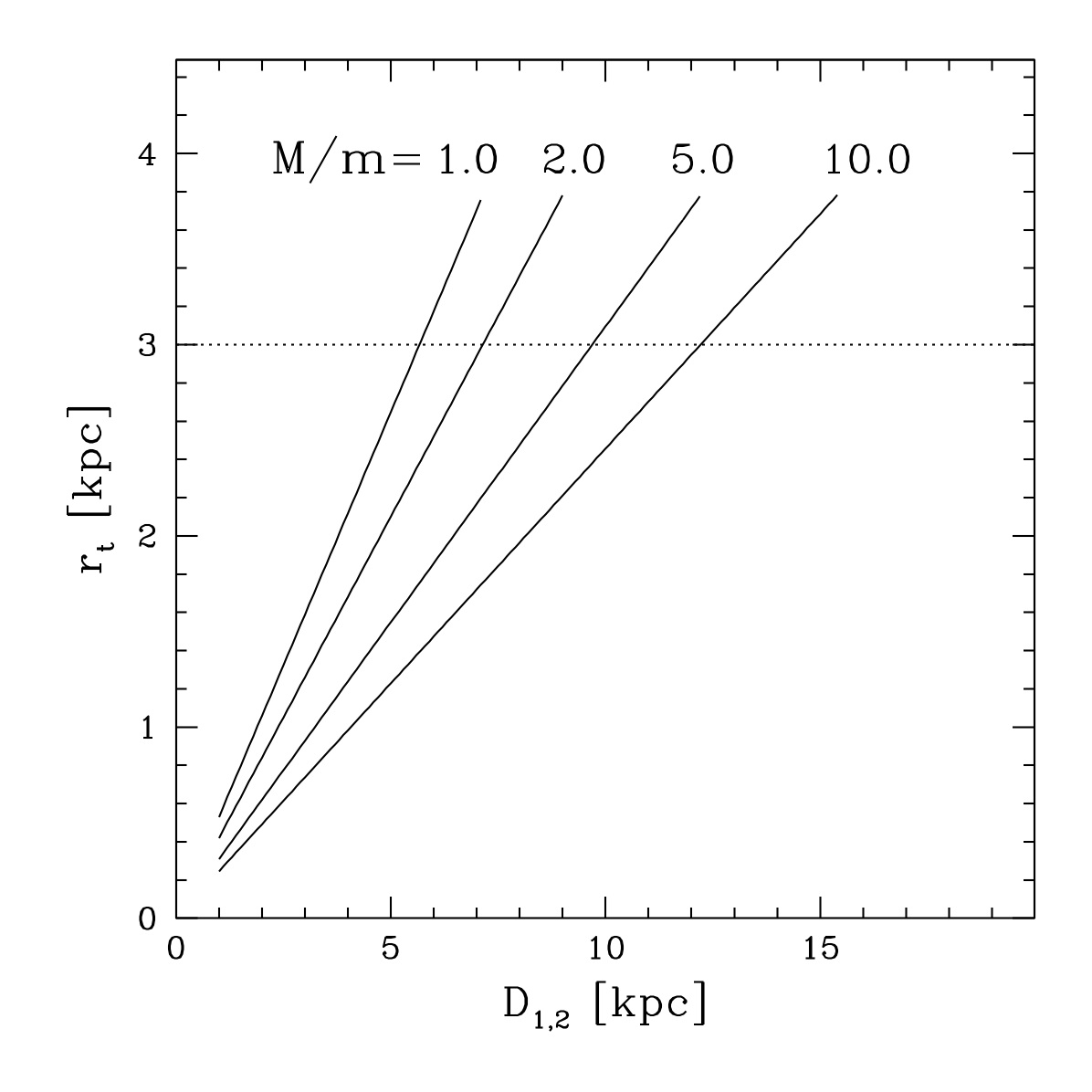}
     \caption{Tidal radius as a function of mutual distance ($D_{1,2}$ between two galaxies of various mass ratios (M/m), from Eq.~\ref{RT}. The dotted line mark the approximate tidal radius of Sex~A and Sex~B, the range of M/m considered in the figure corresponds to the mass ratios between Sex~A and Sex~B, and between Sex~A/Sex~B and NGC~3109.    }
        \label{combT}
    \end{figure}


Assuming $\simeq 3$~kpc as the approximate truncation radius for this galaxy (from Fig.~\ref{prof}) we show in Fig.~\ref{combT} what is the perigalactic distance at which such a limit can be set by tides for various values of the mass ratio of the interacting galaxies. In the range of mass ratios characteristic of an encounter with Sex~B ($M/m=1-2$) a passage in the range 5~kpc$\la D_{1,2}\la$ 8~kpc is required, while for encounters with NGC~3109 a passage at  10~kpc$\la D_{1,2}\la$ 13~kpc will be sufficient. We note that Sex~A is $\simeq 250$~kpc apart from Sex~B and $\simeq 508$~kpc from NGC~3109 \citep[see][]{filam}, hence, at present it is essentially undisturbed by the tides of its closest companions. 
For M/m$\ge 1000$, corresponding to the mass ratio between Sex~A/Sex~B and the MW, the $r_t=3$~kpc line is matched at $D_{1,2}\ge 57$~kpc, similar to the peri-galactic distance of the Magellanic Clouds, according to the most recent analysis \citep{kalli}.

   \begin{figure}
   \centering
   \includegraphics[width=\columnwidth]{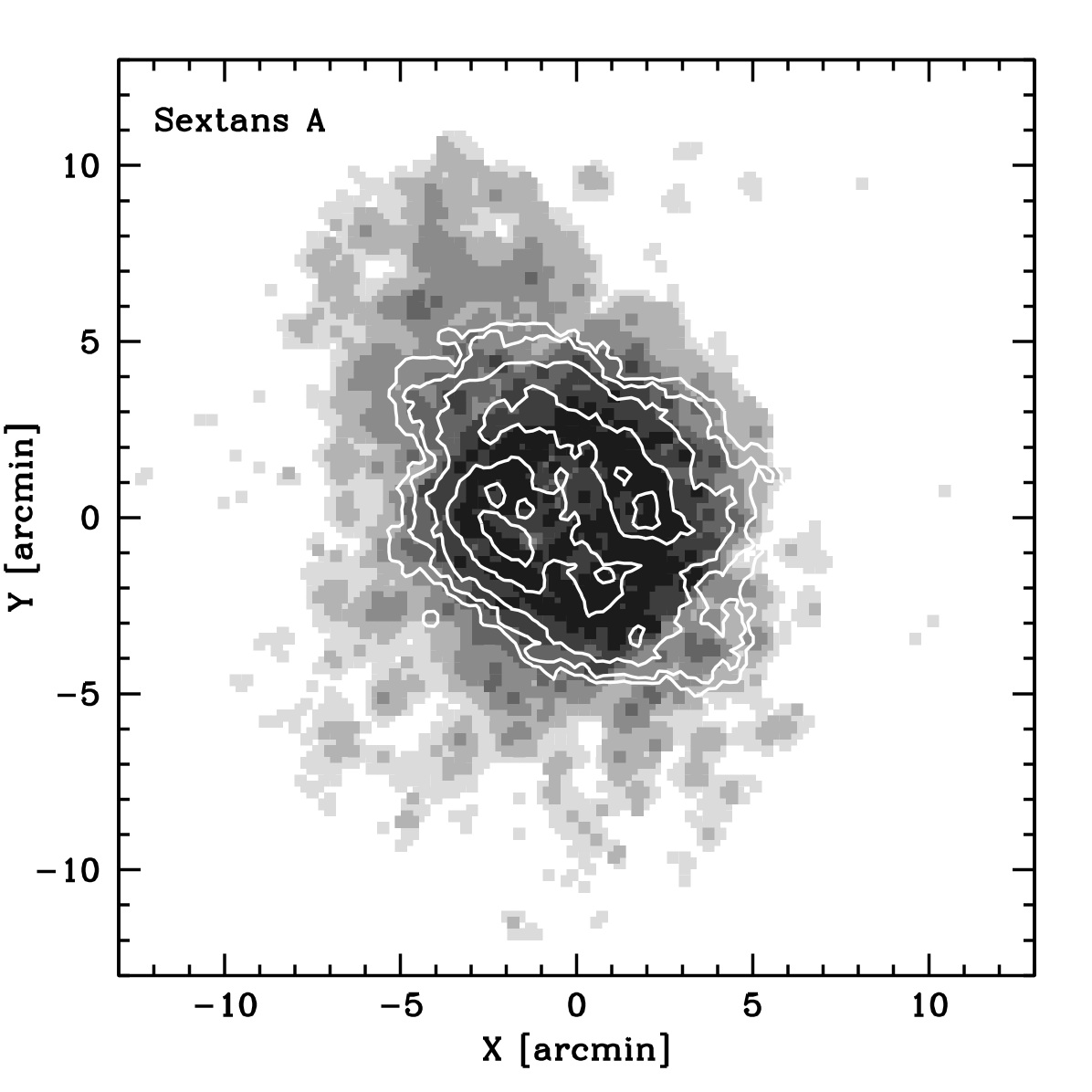}
   \includegraphics[width=\columnwidth]{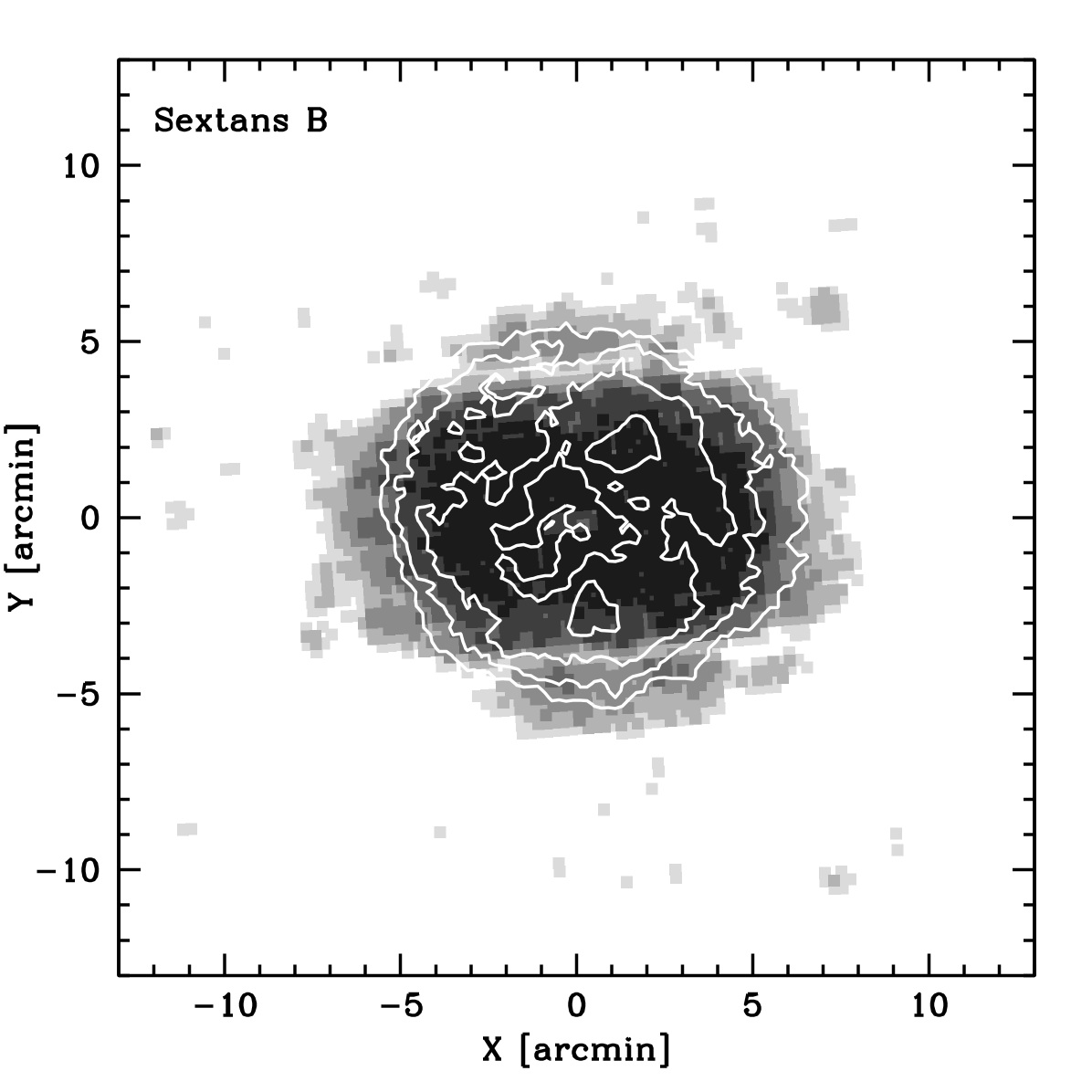}
     \caption{\HI~ surface density contours (white) derived from \citet{ott} maps are superposed to the stellar density maps obtained with the matched filter thechnique (greyscale). 
 Contour levels are 3, 9, 27, 81, 240 and 720 $\times 10^{19}$~cm$^{-2}$.  The levels of grey correspond to density of 3, 5, 10, 20, 40 and 80$\sigma$ above the background, from the lightest to the darkest tone of grey.}
        \label{HImaps}
    \end{figure}


\section{Neutral Hydrogen}
\label{HI}

We have analyzed the distribution and kinematics of the neutral Hydrogen using the publicly available deep observations of Sex~A and Sex~B provided by the  
VLA-ANGST \citep{ott}\footnote{\tiny\tt https://science.nrao.edu/science/surveys/vla-angst/} survey.

\begin{figure}
   \centering
   \includegraphics[width=\columnwidth]{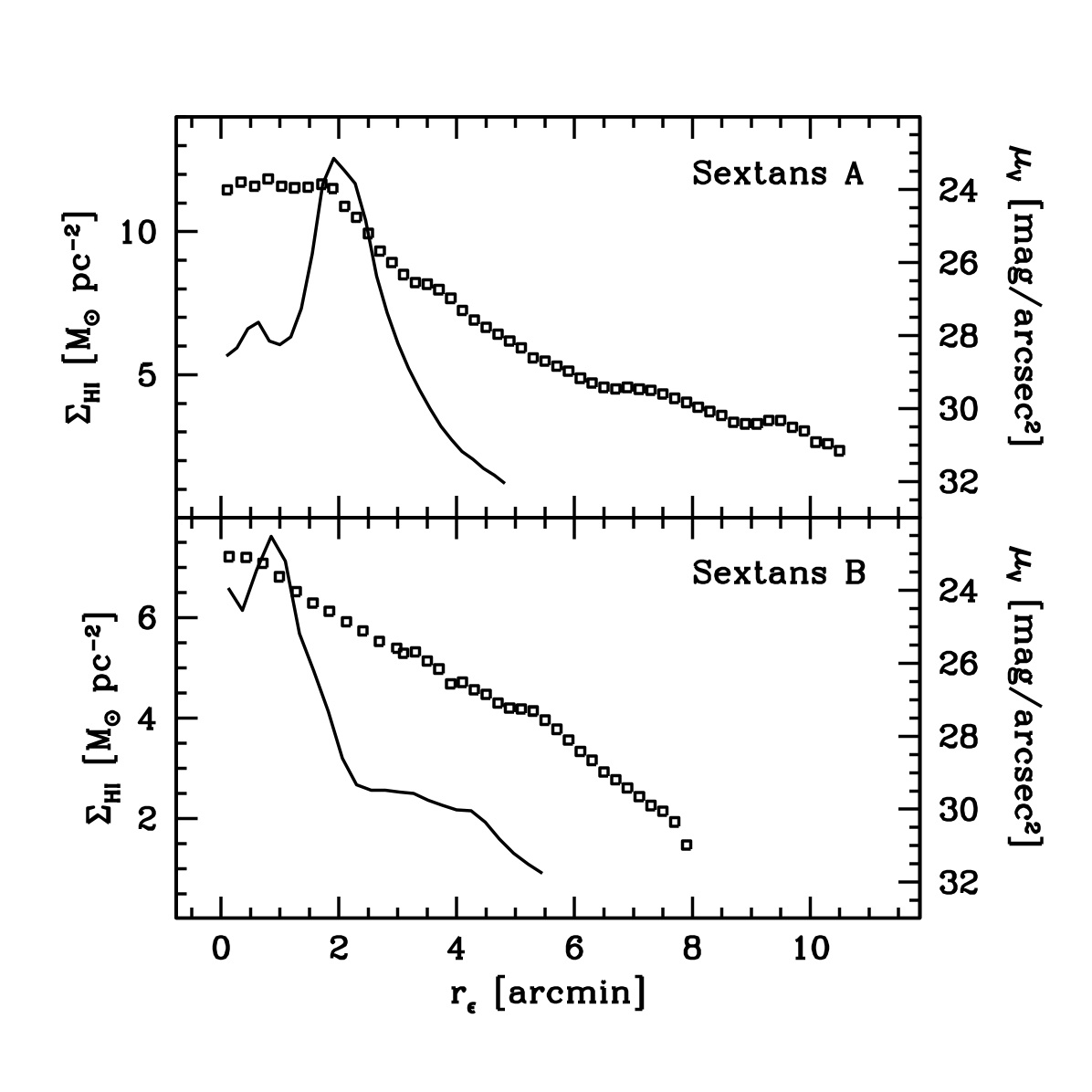}
     \caption{\HI~ surface density profiles (continuous lines, scale on the left vertical axis) compared to surface brightness profiles (empty squares, scale on the right vertical axis), for Sextans~A (upper panel) and Sextans~B (lower panel).}
        \label{HIprof}
    \end{figure}

In Fig.~\ref{HImaps} we superpose the \HI~ surface density contours to the stellar surface density maps we obtained from star counts with the matched filter technique (see Pap-I).
There are a few interesting conclusions that can be drawn from this simple comparison:

\begin{itemize}

\item In both galaxies the neutral Hydrogen component is comparably/{\em less} extended than the distribution of the stars (see also Fig.~\ref{HIprof}). This was already noted to happen in VV124 (Pap-I) but that galaxy have a ratio of \HI~ mass to stellar mass $M_{\HI}/M_*\simeq 0.1$, while Sex~A and Sex~B have $M_{\HI}/M_*\ga 1.0$ (adopting ${\frac{M}{L_V}}=1.0$, in solar units, for the stellar component), i.e. they are {\em much} more gas-rich than VV124. It must be recalled that the truncation of the \HI~ distribution does not necessarily imply that no bound gas is present at larger radii, since extended ionized gas of very low surface brightness is expected to surround the \HI~ discs of spirals and irregulars \citep{maloney,corbe}.

\item The outer contours of the gas and the star surface density distributions have remarkably similar shapes in Sex~B. The same is true for the main body of Sex~A. In general the overall large-scale correlation between the gas and the stars is pretty high. This was far from obvious, since the stellar maps are based on old-intermediate age tracers. On the other hand, the correlation is very strong between the highest density peaks of the \HI~ and the youngest stars and associations, as expected.

\item The \HI~ distribution in Sex~A seems to end near the $r_{\epsilon}\simeq 6\arcmin$ break in the SB profile, marking of the emergence of the wide elongated halo surrounding the galaxy. This supports the interpretation of this structure as a tidal tail. A tidal interaction would strain and strip stars and gas in equal measure, but the stripped gas would be easily dissipated by ram pressure 
\citep[also induced by the corona of gas ejected from the group members, see][]{shen} while the stars can remain partially bound, or unbound but on orbits very similar to the main body of the galaxy, for a long time. The outer \HI~ warp is also compatible with some tidal disturbance in the past.

\end{itemize}

In Fig.~\ref{HIprof} we compare the \HI~ and SB profiles. It is interesting to note that the inner flattening of the SB profile in Sex~A corresponds to a central drop in the \HI~ column density, possibly suggesting that the recent strong star formation activity may have a role in the formation of a constant SB core \citep[see, e.g.,][and references therein]{core}. A similar correspondence, on much smaller scales, may be present also in the profile of Sex~B, for $r_{\epsilon}\la 1.0\arcmin$.

We have performed tilted ring fitting of the velocity fields and build some 3D models of the data-cubes using the standard GIPSY software \citep[see e.g.,][]{batta}.
We found that  in both galaxies the kinematics is clearly dominated by rotation although some non-circular motions may be present. 
Both Sex~A and Sex~B show nearly solid-body rotation with some indication of flatting in the outer parts.
The kinematic position angles of the discs are well determined within a few degrees and are $96\degr$ and $61\degr$ for Sex A and B, respectively.
For Sex A this refers to the inner disc while in the outer parts there are signs of a warp and a change in the position angle towards lower values.

Unfortunately, the inclination angles are very difficult to constrain using the HI data-cubes as it often happens for solid body rotation.
Acceptable fits can be obtained with a wide range of inclinations provided that the thickness of the disc changes accordingly.
For instance, for Sex~B acceptable models are obtained with inclinations spanning from $20\degr$ to almost $80\degr$.
In the case of $i=20\degr$ , the disc can be quite thin with (exponential) scale-height lower than $300$ pc, while if one assumes 80 degrees the required scale-height is about 1 kpc.
If we assume an intermediate value of i=40 degrees, the rotation at the last measured point is $v_{last}\sim50$ km~s$^{-1}$.

Also in the case of Sex~A, the inclination is rather unconstrained by the HI kinematics although it is clearly different from zero given the rotation pattern observed. The overall morphology of the HI and of the stellar body suggests a relatively low inclination.
If we assume $i=20\degr$ we obtain a $v_{last}\sim50$ km~s$^{-1}$ also for this galaxy.

Under these assumptions, the dynamical masses estimated at the last measured points of the rotation curves would be of about $1\times 10^{9}~M_{\sun}$, implying a mass-to-light ratio $\frac{M}{L_V}\sim 25$ and a dark-to-barionic mass ratio of $\sim 10$.
We stress that these values are very uncertain given the low inclinations of the galaxies, for instance an inclination as low as $i=20\degr$ for Sextans~B would lead to a dynamical mass four times larger.

\section{Summary and Discussion}
\label{disc}

We present new deep wide-field photometry, and a re-analysis of existing \HI~ data for the two isolated dwarf irregular galaxies Sextans~A and Sextans~B, members of the NGC~3109 association of dwarfs \citep{tully06,filam}. The main results of the present study can be summarized as follows:

\begin{itemize}

\item In both galaxies there are obvious population gradients on a global scale. As generally observed
in other dwarfs, younger stars are confined to the innermost regions while older and older stars have more and more extended distributions. The mean metallicity of the old stellar component is low  ($\langle [Fe/H]\rangle\la -1.4$) with no obvious metallicity gradient. We provide the first estimate of the mean metallicity for the oldest stellar population of Sex~B. Using the color distribution of bright RGB stars from HST-ACS photometry we obtain $\langle [Fe/H]\rangle_{SexB}=-1.6$.

\item One candidate compact star cluster has been identified in Sex~A and two in Sex~B, in addition to the already confirmed C1 in Sex~B \citep{shaclus}. 

\item We extended the existing SB profiles from surface photometry (reaching $\mu_V\sim 27.5$~mag/arcsec$^2$) with profiles from star counts reaching $\mu_V\sim 31$~mag/arcsec$^2$. Both galaxies display SB profiles that are not adequately reproduced by a single \citet{sersic} model. The profile of Sex~B shows a down-bending at $r_{\epsilon}\simeq 5\arcmin$, while that of Sex~A presents a flat central region (for $r_{\epsilon}\simeq 1.8\arcmin$, strongly correlated with a drop in the \HI~ surface density profile), and an upward bending at $r_{\epsilon}\simeq 6\arcmin$.

\item We show that the stellar body of the two galaxies is much more extended that what previously traced with surface photometry. In particular we can trace the profiles of Sex~A and Sex~B out to $\sim 4$~kpc and $\sim 3.5$~kpc, respectively (w.r.t $\la 2$~kpc). In both cases the stellar body appears to be more extended than the \HI~ disc.

\item The kinematics of the \HI~ are dominated by a significant solid-body rotation. The large uncertainties on the inclinations of the discs prevents an accurate determination of the maximum rotational velocity. Under conservative assumptions we find $v_{last}\simeq 50$~km~s$^{-1}$ for both galaxies, corresponding to a dynamical mass of $\sim 10^{9}~M_{\sun}$.

\item The stellar density map of Sex~A shows that the roundish main body of the galaxy, nicely matched by  the \HI~ disc in position and shape, is surrounded by a flatter and elongated LSB halo of stars. This structure appears to display a broad {\tt S}-shape and its onset corresponds to the outer up-bending change in slope of the SB profile. Both features are typical of tidal tails \citep{katy,pena_t,munoz} and also the structure and kinematics of the \HI~, with a warp in correspondence with the position of the tail, are fully compatible with the the hypothesis of a tidal origin for the observed structure. It is very interesting to note that
a similar feature has been recently identified in another member of the same association Sex~A belongs to, the Antlia dIrr, and it has been interpreted as a tidal tail due to a past interaction with NGC~3109 \citep{penny}.

\item Basic considerations on the size and masses of the involved galaxies suggests that both an interaction with another member of the NGC~3109 association or with the Milky Way are plausible as possible origin for the putative tidal tail of Sex~A. In any case, the comparison between the relative distances and velocities indicates that these interactions should have happened several Gyrs ago (see below). This may appear as a problem for the tidal interpretation, since tidal tails are not expected to be very long-lived. However it must be recalled that, in fact, the very low SB portions of the tails can persist for long times \citep[see][e.g., their Fig.~2]{pena_t}, and that, in general, available theoretical studies have considered the case of dwarfs orbiting a main galaxy, while in the present case it is likely that we are dealing with a single close encounter (possibly between dwarfs) followed by an evolution in (progressively deeper) isolation. Dedicated N-body simulations are probably needed to adequately explore this specific case.


\end{itemize}

The last point should be considered in the context of the past history of the NGC~3109 association.
\citet{teys12} compared the observed distances and velocities of LG dwarfs with those of state-of-the-art N-body simulations, to deduce their past orbital histories. In their analyses, from a sample of 39 relatively isolated dwarfs, they identified 10 galaxies as very likely {\em back-splashers}, i.e. stellar system that have passed within 0.5 virial radii of a dominant member of the LG (MW or M31) in the past.
Among these ten, they list NGC~3109, Antlia, Sex~A, and Sex~B, i.e. all the members of the NGC~3109 association that were known at the time of the Teyssier et al.'s analysis\footnote{The newly discovered Leo~P galaxy \citep{giova} is likely a new member of the association, see \citealt{filam}, and references therein.}. In their independent attempt of reconstruction of the past path of all the LG galaxies, also \citet{shaya} envisage a passage at $\sim 25$~kpc from the MW of most of the members of the association, occurred $\sim 7$~Gyr ago. 

The hypothesis of a close encounter of the NGC~3109 association with the MW provides a natural scenario to explain the recent finding that all the members of the association lie within $\sim 100$~kpc (r.m.s) of a line in space that is $\sim 1.1$~Mpc long, from one extreme (NGC~3109) to the other (Leo~P), and display a gradient between velocity in the LG reference frame ($V_{LG}$) and distance along the line 
of 43~km~s$^{-1}$~Mpc$^{-1}$, with a r.m.s. of just 17~km~s$^{-1}$ \citep{filam}. It is interesting to note that the timescale for the association to reach the current distance from the MW (the ratio between this distance and the mean velocity of the association in the Galactic reference frame, $\sim 7$~Gyr) is similar to the timescale to scatter by 100~kpc from the line by the effect of random motions of 17~km~s$^{-1}$ ($\sim 6$~Gyr). This suggests that, if indeed the association had a close encounter with the MW a few Gyrs ago, at that epoch its members were very close together and/or tightly clustered in a filamentary configuration, and {\em dynamically cold}, otherwise they would be much more scattered in space at the present epoch. On the other hand, no obvious signature of this interaction can be identified in the SFHs of the members of the association presented by \citet[][except perhaps for Antlia, whose star formation rate peaks $\sim 6$~Gyr ago]{weisz}. However, SF enhancements on scales shorter than $\sim 1$~Gyr can be easily wiped out by the low temporal resolution ($\sim 3-4$~Gyr) of the reconstructed SFHs in that range of look-back times.

Given all the above arguments, in a Newtonian dynamics and $\Lambda$- cold dark matter framework \citep{bark}, it is plausible to think that the members of the NGC~3109 association, after decoupling from the Hubble flow, were originally falling toward
the Milky Way along one of the thin and cold cosmological filaments that are predicted by the current paradigm of galaxy formation \citep{dekel,lovell}\footnote{See \citet{filam}, for alternative hypotheses.}. The close encounter with the Galaxy tidally stretched the dwarfs into their present extremely elongated and dynamically cold configuration, also producing (or enhancing) the observed velocity gradient, and finally pulling all of them out of the LG \citep{shaya}. The peri-Galactic passage was not only the time when the tidal strain of the MW was maximum on each individual member of the association, but also the epoch when the members were most closely packed together, maximizing the effects and the probability of dwarf-to-dwarf interactions. It seems likely that the tidal tails of Antlia and Sex~A, as well as the warp and sub-structures observed in NGC~3109 \citep{hida,blok,dal09} were produced during that (relatively) short epoch of multiple interactions.

\begin{acknowledgements}

We acknowledge the support from the LBT-Italian Coordination Facility for the
execution of observations, data distribution and reduction.

M.B and F.F. acknowledge the financial support from PRIN MIUR 2010-2011 project ``The
Chemical and Dynamical Evolution of the Milky Way and Local Group Galaxies'',
prot. 2010LY5N2T. 

We are grateful to D. Hunter for providing the surface photometry
of Sex~A and Sex~B in electronic form. 
We thank the VLA-ANGST project for making their HI datacubes and maps publicly
available. We also thank Juergen Ott for helpful information on the data reduction.
We are grateful to an anonymous referee
for useful comment and suggestions that improved the overall quality and focus
of the paper.

This research has made use of the SIMBAD database, operated at CDS, Strasbourg, France.
This research has made use of the NASA/IPAC Extragalactic Database (NED) which is operated by the Jet Propulsion Laboratory, California Institute of Technology, under contract with the National Aeronautics and Space Administration.
This research make use of SDSS data. 
Funding for the SDSS and SDSS-II has been provided by the Alfred P. Sloan Foundation, the 
Participating Institutions, the National Science Foundation, the U.S. Department of Energy, the National Aeronautics and Space Administration, the Japanese Monbukagakusho, the Max Planck Society, and the Higher Education Funding Council for England. The SDSS Web Site is http:\/\/www.sdss.org\/.
The SDSS is managed by the Astrophysical Research Consortium for the Participating Institutions. The Participating Institutions are the American Museum of Natural History, Astrophysical Institute Potsdam, University of Basel, University of Cambridge, Case Western Reserve University, University of Chicago, Drexel University, Fermilab, the Institute for Advanced Study, the Japan Participation Group, Johns Hopkins University, the Joint Institute for Nuclear Astrophysics, the Kavli Institute for Particle Astrophysics and Cosmology, the Korean Scientist Group, the Chinese Academy of Sciences (LAMOST), Los Alamos National Laboratory, the Max-Planck-Institute for Astronomy (MPIA), the Max-Planck-Institute for Astrophysics (MPA), New Mexico State University, Ohio State University, University of Pittsburgh, University of Portsmouth, Princeton University, the United States Naval Observatory, and the University of Washington.

\end{acknowledgements}

\bibliographystyle{apj}

\begin{thebibliography}{999}


\bibitem[Ahn et al.(2012)]{dr9} 
         Ahn, C.P., Alexandroff, R., Allende Prieto, C., et al. 2012, \apjs, 
	 203, 21                     
\bibitem[Barkana \& Loeb(2001)]{bark}
         Barkana, R., \& Loeb, A., 2001, Ph. Rep., 349, 125 
\bibitem[Barnes \& de Blok(2001)]{blok}
         Barnes, D.G., \& de Blok, W.J.G., 2001, \aj, 122, 825
\bibitem[Battaglia et al.(2006)]{batta} 
         Battaglia, G., Fraternali, F., Oosterloo, T., Sancisi, R.,\aap, 447, 49
\bibitem[Beccari et al.(2010)]{comple}
         Beccari, G., Pasquato, M., de Marchi, G., Dalessandro, E., Trenti, M.,
	 \& Gill, M., \apj, 713, 194
\bibitem[Bellazzini et al.(2002a)]{lf}
         Bellazzini, M., Fusi Pecci, F., Messineo, M., Monaco, L., 
	 \& Rood, R. T. 2002a, \aj, 123, 1509 
\bibitem[Bellazzini et al.(2002b)]{umi}	 
         Bellazzini M., Ferraro F. R., Origlia L., Pancino E., Monaco L., Oliva
	 E., 2002b, \aj, 124, 3222
\bibitem[Bellazzini, Ferraro \& Pancino(2001)]{tip1}
         Bellazzini, M., Ferraro, F.R., \& Pancino, E., 2001, \apj, 556, 635
\bibitem[Bellazzini et al.(2004a)]{tip2}	
         Bellazzini, M., Ferraro, F. R., Sollima, A., Pancino, E., \& 
	     Origlia, L., 2004a, \aap, 424, 199
\bibitem[Bellazzini et al.(2004b)]{tidal}
         Bellazzini, M., 2004b,\mnras, 347, 119	 
\bibitem[Bellazzini(2008)]{cefatip}
         Bellazzini, M., 2008. Mem. SAIt, 79, 440 (B08)
\bibitem[Bellazzini et al.(2011a)]{pap1}
         Bellazzini, M., Beccari, G., Oosterloo, T.A., et al., 2011a, \aap, 527, A58	 
\bibitem[Bellazzini et al.(2011b)]{vvhst}
         Bellazzini, M., Perina, S., Galleti, S., Oosterloo, T.A., 2011b, \aap, 533, 37
\bibitem[Bellazzini et al.(2013)]{filam}
         Bellazzini, M., Oosterloo, T., Fraternali. F., Beccari, G., 2013, 
	 \aap, 559, L11
\bibitem[Bertin \& Arnouts(1996)]{sex}
         Bertin, E., \& Arnouts, S., 1996, \aaps, 117, 393
\bibitem[Ciotti \& Bertin(1999)]{ciotti}
         Ciotti, L.,  \& Bertin, G., 1999, \aap, 352, 447
\bibitem[Clem et al.(2008)]{clem} 
         Clem, J.L., Vandenberg, D.A., Stetson, P.B., 2008, \aj, 135, 628
\bibitem[Carretta \& Gratton(1997)]{cg97}
         Carretta, E., \& Gratton, R., 1997, \aaps, 121, 95
\bibitem[Carretta et al.(2009)]{euge}
         Carretta, E., Bragaglia, A., Gratton, R.G., D'Orazi, V., 
	 \& Lucatello, S., 2009, \aap, 508, 695	
\bibitem[Corbelli \& Salpeter(1993)]{corbe}
         Corbelli, E., \& Salpeter, E.E., 1993, \apj, 419, 104
\bibitem[Cotton et al.(1999)]{cotton}
         Cotton, W.D., Condon, J.J., Arbizzani, E., 1999, \apjs, 125, 409
\bibitem[Dalcanton et al.(2009)]{dal09}
         Dalcanton, J.J., Williams, B.F., Seth, A.C., et al., 2009, \apjs, 183, 67
\bibitem[Dekel et al.(2009)]{dekel}
         Dekel, A., Birnboim, Y., Engel, G., et al., 2009, Nature, 457, 451
\bibitem[Dohm-Palmer et al.(1997a)]{dohm_cmd}
         Dohm-Palmer, R.C., Skillman, E.D., Saha, A., et al., 1997a, \aj, 114,
	 2514
\bibitem[Dohm-Palmer et al.(1997b)]{dohm_sfh}
         Dohm-Palmer, R.C., Skillman, E.D., Saha, A., et al., 1997b, \aj, 114,
	 2527
\bibitem[Dolphin et al.(2003)]{dolph}
         Dolphin, A.E., Saha, A., Skillman, E.D., et al., 2003, \aj, 126, 187	 
\bibitem[Federici et al.(2007)]{lucky}
         Federici, L., Bellazzini, M., Galleti, S., Fusi Pecci, F., Buzzoni, A.,
	 \& Parmeggiani, P., 2007, \aap, 473, 429
\bibitem[Ferguson et al.(2007)]{ann_m33}
         Ferguson, A.M.N, Irwin, M., Chapman, S., Ibata, R., Lewis, G.F., Tanvir, N., 
	 2007, in Island Universes, Astroph. Sp. Sci. Proc., Springer, p. 239
\bibitem[Gallart(2008)]{lcid} 
         Gallart, C.,  2008, ASP Conf. Ser., 390, 278 
\bibitem[Galleti et al.(2007)]{silvr}	
	     Galleti, S., Bellazzini, M., Federici, L., Buzzoni, A., 
	     \& Fusi Pecci, F., 2007, \aap, 471, 127		      
\bibitem[Georgiev et al.(2010)]{geo}
         Georgiev, I.Y., Puzia, T.H., Goudfrooij, P., Hilker, M., 2010, \mnras,
	 406, 1967
\bibitem[Giallongo et al.(2008)]{lbc} 
         Giallongo, E., Ragazzoni, R., Grazian, A., 2008, \aap 482, 349	
\bibitem[Giovanelli et al.(2013)]{giova}
         Giovanelli, R., Haynes, M.P., Adams, E., 2013, \aj, 146, 15
\bibitem[Girardi et al.(2004)]{gira}		       
         Girardi, L., Grebel, E.K., Odenkirchen, M., Chiosi, C., 204, \aap, 422,
	 205
\bibitem[Johnston et al.(1999)]{katy}
         Johnston, K.V., Sigurdsson, S., \& Hernquist, L., 1999, \mnras, 302, 771
\bibitem[Harbeck et al.(2001)]{harbeck} 
         Harbeck D. et al., 2001, \aj, 122, 3092
\bibitem[Harris(1996)]{harris}
         Harris, W.E.,1996, \aj, 112, 1487
\bibitem[Hidalgo et al.(2008)]{hida}
         Hidalgo, S.L., Aparicio, A., Gallart, C., 2008, \aj, 136, 2332 
\bibitem[Hunter \& Elmegreen(2006)]{HE06} 
         Hunter, D.A., \& Elmegreen, B.G., 2006, \apj, 162, 49 (HE06)
\bibitem[Huxor et al.(2013)]{N6822}
         Huxor, A.P., Ferguson, A.M.N., Veljanoski, J., Mackey, A.D., \& Tanvir,
	 N.R., 2013, \mnras, 429, 1039
\bibitem[Huxor et al.(2005)]{hux}
         Huxor, A.P., Tanvir, N.R., Irwin, M.J., et al., 2005, \mnras, 360, 1007
\bibitem[Hwang et al.(2011)]{hwang}
         Hwang, N., Lee, M.G., Lee, J.C., et al., 2011, \apj, 738, 58
\bibitem[Kallivayalil et al. (2013)]{kalli}
         Kallivayalil, N., van der Marel, R.P., Besla, G., Anderson, J., Alcock, C., 
	 2013, \apj, 764, 161
\bibitem[Kazantzidis et al.(2010)]{kaza}
         Kazantzidis, S., Lokas, E.L., Callegari, S., Mayer, L., \& Moustakas,
	 L.A., \apj, submitted (arXiv:1009.2499) 	 				       
\bibitem[Kaufer et al.(2004)]{kaufer}
         Kaufer, A., Venn, K.A., Tolstoy, E., Pinte, C., \& Kudritzki, R.-P., 
	 \aj, 127, 2723
\bibitem[Keenan(1981a)]{keen_a}
         Keenan, D.W., 1981a, \aap, 71, 245
\bibitem[Keenan(1981b)]{keen_b}
         Keenan, D.W., 1981b, \aap, 95, 340
\bibitem[Kirby et al. (2012)]{kvo124}
         Kirby, E.N., Cohen, J.G., Bellazzini, M., 2012, \apj, 751, 46
\bibitem[Kirby et al. (2013)]{kv124}
         Kirby, E.N., Cohen, J.G., Bellazzini, M., 2013, \apj, 768, 96
\bibitem[Klimentowski et al.(2009)]{klim_tid}
         Klimentowski, J., Lokas, E.L., Kazantzidis, S., Mayer, L., Mamon, G.A.,
	     Prada, F., 2009, \mnras, 400, 2162 
\bibitem[Kniazev et al.(2005)]{kniazev}
         Kniazev, A.Y., Grebel, E.K., Pustilnik, S.A., Pramskij, A.G., \&
	 Zucker, D.B., 2005, \aj, 130, 1558
\bibitem[Lauer(1985)]{xvista}	 
	 Lauer, T. R. 1985, \apjs, 57, 473
\bibitem[Lovell et al.(2011)]{lovell}
         Lovell, M.R., Eke, R.E., Frenk, C.S., \& Jenkins, A., 2011, \mnras,
	 413, 3013
\bibitem[Magrini et al.(2005)]{magrini}
         Magrini, L., Leisy, P., Corradi, R.L.M., Perinotto, M., Mampaso, A., \&
	 V\`ilchez, J.M., 2005, \aap, 443, 115
\bibitem[Maloney(1992)]{maloney}
         Maloney, P., 1992, \apj, 398, L89
\bibitem[Massey et al.(2007)]{massey}
         Massey, P., Olsen, K.A.G., Hodge, P.W., Jacoby, G.H., McNeill, R.T.,
	 Smith, R.C., Strong, S.B., 2007, \aj, 133, 2393
\bibitem[Mateo(1998)]{mateo}
         Mateo, M., 1998, \araa, 36, 435
\bibitem[Mayer et al.(2007)]{lucionat}
         Mayer, L., Kazantzidis, S., Mastropietro, C., \&  Wadsley, J., 2007, Nature, 445, 738
\bibitem[McConnachie(2012)]{mcc}
         McConnachie, A., 2012, \aj, 144, 4 (M12)
\bibitem[Monelli et al.(2010a)]{monel_cet}
         Monelli, M., Hidalgo, S.L., Stetson, P.B., et al., 2010a, \apj, 720, 1225
\bibitem[Monelli et al.(2010b)]{monel_tuc}
         Monelli, M., Gallart, C., Hidalgo, S.L., et al., 2010b, \apj, 722, 1864
\bibitem[Mu\~noz, Majewski, \& Johnston(2008)]{munoz}
         Mu\~noz, R.R., Majewski, S.R., \& Johnston, K.V., 2008, \apj, 679, 346
\bibitem[Ott et al.(2012)]{ott}
         Ott, J., Stilp, A.M., Warren, S.R., et al, \aj, 144, 123
\bibitem[Peacock et al.(2011)]{pea}
         Peacock, M.A., Zepf, S.E., Maccarone, T.J., \& Kundu, A., 2011, \apj, 737, 5
\bibitem[Pedreros \& Gallart(2002)]{pg02}
         Pedreros, M.H., Gallart, C., 2002, Rev. Mex. AA, 14, 73
\bibitem[Pe\~narrubia, Navarro \& McConnachie(2008)]{pena}
         Pe\~narrubia, J., Navarro, J.F., \& McConnachie, A.W., 2008, \apj, 673, 226
\bibitem[Pe\~narrubia et al.(2009)]{pena_t}        
         Pe\~narrubia, J., Navarro, J.F., McConnachie, A.W., Martin, N.F., 
	 2008, \apj, 698, 222
\bibitem[Penny et al.(2012)]{penny}
         Penny, S.J., Pimbblet, K.A., Conselice, C.J., Brown, M.J.I., 
	 Gr\"utzbauch, R., Gloyd, D.J.E., 2012, \apj, 758, L32
\bibitem[Pohlen \& Trujillo(2006)]{pohlen}
         Pohlen, M., \& Trujillo, I., 2006, \aap, 454, 759
\bibitem[Rockosi et al.(2002)]{connie}
         Rockosi, C.M., Odenkirchen, M., Grebel, E.K., 2002, \aj, 124, 349
\bibitem[Sakai et al.(1996)]{sakai}
         Sakai, S., Madore, B.F., Freedman, W.L., 1996, \apj, 461, 713
\bibitem[Sanna et al.(2010)]{sanna}
         Sanna, N., Bono, G., Stetson, P.B., et al., 2010, \apj, 722, L244
\bibitem[Sawala et al.(2010)]{saw10}
         Sawala, T., Scannapieco, C., Maio, U., White, S., 2010, \mnras, 402, 1599
\bibitem[Sawala et al.(2012)]{saw12}
         Sawala, T., Scannapieco, C., White, S., 2012, \mnras, 420, 1714
\bibitem[Schlafly \& Finkbeiner(2011)]{schlaf}
         Schlafly, E., \& Finkbeiner, D.P., 2011, \apj, 737, 103
\bibitem[Schlegel et~al.(1998)]{ebv}
         Schlegel, D.~J., Finkbeiner, D.~P., \& Davis, M. 1998, \apj, 500, 525
\bibitem[S\'ersic(1968)]{sersic}
         S\'ersic, J.L., 1968, Atlas de Galaxias Australes, Cordoba, Argent.:
	 Obs. Astron.
\bibitem[Sharina et al.(2007)]{shaclus}
         Sharina, M.E., Puzia, T.H., Krylatyh, A.S., 2007, Astroph. Bull., 63,
	 209
\bibitem[Sharina et al.(2008)]{shaprof}
         Sharina, M.E., Karachentsev, I.D., Dolphin, A.E., et al., 2008, 384,
	 1544	 
\bibitem[Shaya \& Tully(2013)]{shaya}
	 Shaya, E.J., \& Tully, R.B., 2013, \mnras, 436, 2096
\bibitem[Shen et al. (2014)]{shen}
         Shen, S., Madau, P., Conroy, C., Governato, F. Mayer, L., 2014, \apj, 
	 submitted (arXiv:1308.4131)
\bibitem[Stetson(1987)]{daophot}
         Stetson, P.B., 1987, \pasp, 99, 191
\bibitem[Stetson(1994)]{allframe}
         Stetson, P.B., 1994, \pasp, 106, 250
\bibitem[Teyssier et al.(2012)]{teys12}
         Teyssier, M., Johnston, K.V., Kuhlen, M., 2012, \mnras, 426, 1808
\bibitem[Teyssier et al.(2013)]{core}
         Teyssier, R., Pontzen, A., Dubois, Y., Read, J.I., 2013, \mnras, 429, 3068
\bibitem[Tolstoy, Hill \& Tosi et al.(2009)]{tht}
         Tolstoy, E., Hill, V., \& Tosi, M., 2009, \araa, 47, 371
\bibitem[Tucker et al.(2006)]{tuck}
         Tucker, D.L., Kent, S., Richmond, M. W., et al. , 2006, AN, 327, 821	 
\bibitem[Tully et al.(2006)]{tully06}
         Tully, R.B., Rizzi, L., Dolphin, A.E., et al., 2006, \aj, 132, 729 (T06)
\bibitem[van der Kruit(1979)]{kruit}
         van der Kruit, P.C., 1979, \aaps, 38, 15
\bibitem[van den Bergh(1999)]{vdb99}
         van den Bergh, S., 1999, \apj, 517, L97
\bibitem[Vansevi$\check{{\rm c}}$ius et al.(2004)]{vanse}
         Vansevi$\check{{\rm c}}$ius, V., Arimoto, N., Hasegawa, T., et al., 2004, \apj, 611, L93
\bibitem[Weisz et al.(2011)]{weisz}
         Weisz, D.R., Dalcanton, J.J., Williams, B.F., et al., 2011, \apj, 739, 5


\end{thebibliography}


\appendix
\section{Comparison with HST photometry}
\label{quality}

Color Magnitude Diagrams from HST photometry provide the quality benchmark for crowded stellar systems like the distant (but resolved) galaxies we are considering within this project. Since deep HST photometry for Sex~A and Sex~B is publicly available from the ANGST database \citep[][]{dal09}, a direct comparison is worth performing.

\begin{figure}
   \centering
   \includegraphics[width=\columnwidth]{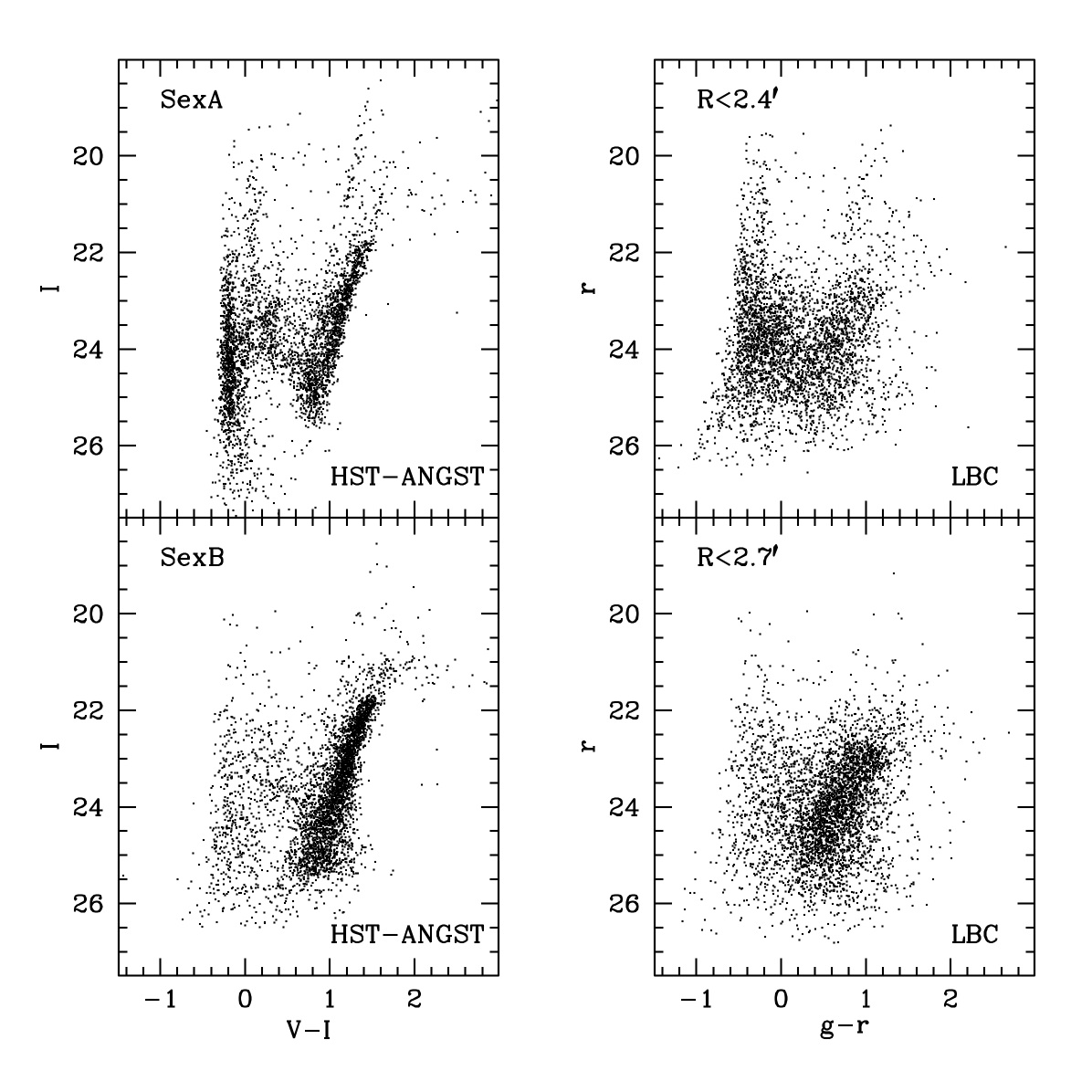}
     \caption{Comparison between HST photometry \citep[left panels, from][]{dal09} and  LBC photometry (right panels)
     for the stars in common between the two datasets.}
        \label{HST}
    \end{figure}

In Fig.~\ref{HST} we show the CMDs obtained from the two datasets for the stars in common. Clearly our ground based photometry cannot rival the top-level quality achievable from space, especially in the highly crowded central regions sampled by the HST data.
Still it is remarkable that all the features that can be identified in the LBC CMDs (e.g., the young MS and the parallel sequence of Blue Loop stars, the RSG plume) have a clear counterpart in the HST CMDs, thus indicating that the precision of
our photometry is sufficient to reliably discriminate stars in different evolutionary phases. 

\section{Compact star clusters}
\label{clus}

It is observed that dIrr galaxies of luminosity similar to Sex~A and Sex~B may host sizable globular clusters systems, with up to $\sim 10$ members \citep[][]{geo}. These clusters can provide very interesting insights into the early formation history of their parent galaxies \citep[see, e.g.][and references therein]{hwang,N6822}. 
The only reference to a systematic search for star clusters in Sex~A and Sex~B available in the literature is by \citet{pg02}; this was a brief contribution to the proceedings of a meeting and no candidate list was provided. On the other hand \citet{shaclus} identified a relatively bright ($M_V=-7.8$) and compact ($r_h\simeq 4$~pc) cluster in Sex~B on HST images where it is marginally resolved into stars
(Sex~B-C1, hereafter). A low resolution spectrum provided further confirmation of the membership of the cluster to Sex~B as well as an estimate of the age ($\simeq 2$~Gyr) and metallicity ([Z/H]=$-1.35\pm 0.3$, \citealt{shaclus}).

\begin{table*}
  \begin{center}
  \caption{Clusters and candidate clusters.}
  \label{clustab}
  \begin{tabular}{lcccccccr}
Name   & $\alpha$ & $\delta$ & g & r & V & FWHM$_r$  & FWHM$_r$/$\langle$FWHM$_{r,stars}\rangle$& notes    \\
\hline
SexA-C1 & 10:10:51.7 & -04:41:33.5   & 20.15   & 19.69 & 19.87 & 2.92$\arcsec$     & 3.9	&  cand. - extended? \\ 
\hline
SexB-C1 & 10:00:04.7 & +05:20:07.5   & 18.16   & 17.87 & 17.98 & 1.04$\arcsec$     & 1.3	&  \citet{shaclus} cluster\\ 
SexB-C2 & 10:00:07.3 & +05:18:00.8   & 18.37   & 17.32 & 17.75 & 1.19$\arcsec$     & 1.5	&  cand. - E0 galaxy?\\ 
SexB-C3 & 09:59:39.7 & +05:16:36.0   & 19.65   & 19.55 & 19.59 & 2.77$\arcsec$     & 3.1	&  cand. - extended?	\\ 
\hline
\end{tabular} 
\tablefoot{Integrated magnitudes are computed within apertures of radius=15~px (corresponding to $3.4\arcsec$), typical uncertainties are $\simeq 0.05$ mag.}
\end{center}
\end{table*}

\begin{figure}
   \centering
   \includegraphics[width=\columnwidth]{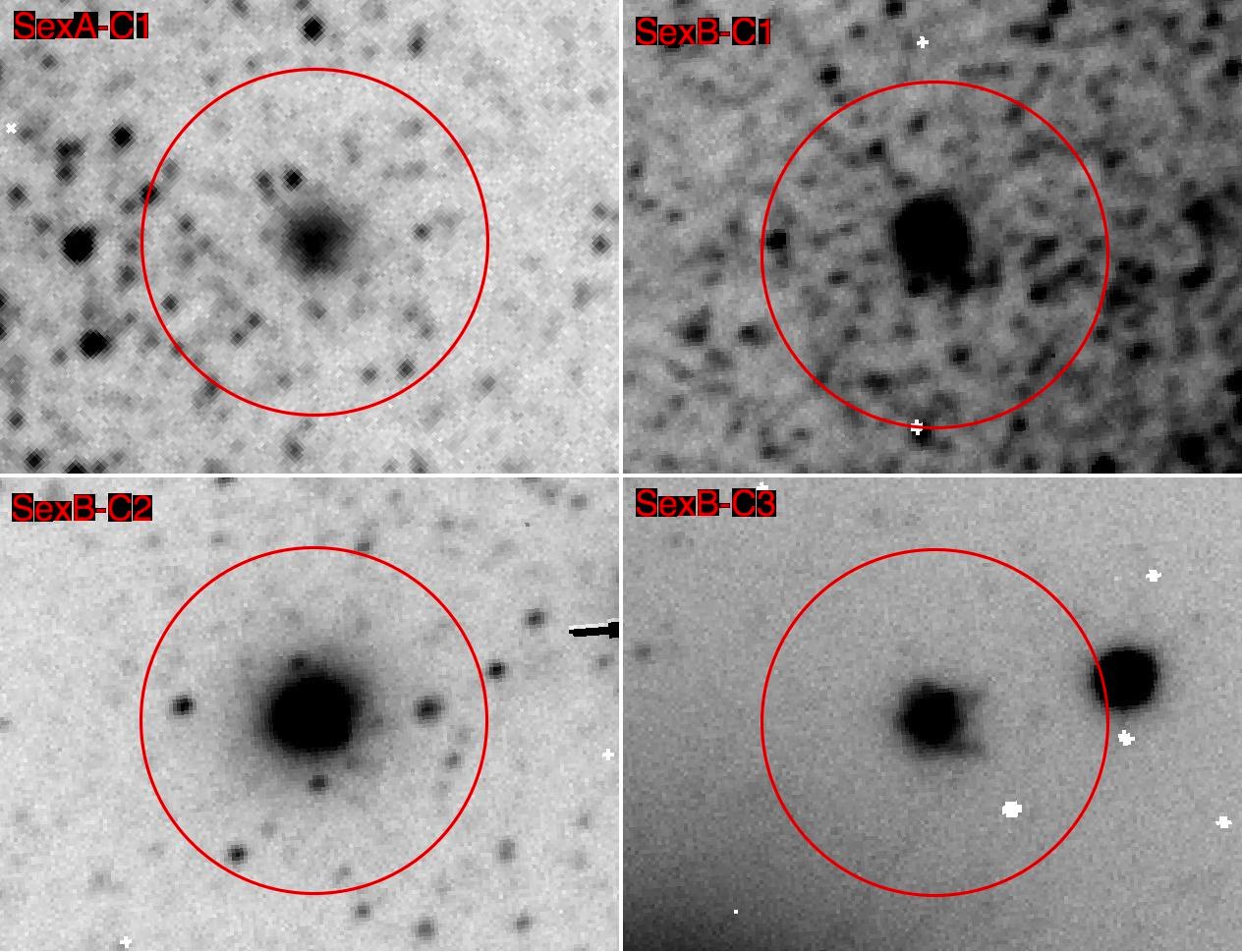}
     \caption{Stamp images of the cluster and candidate clusters listed in Table~\ref{clustab}. The circles around the objects have radius of 10 arcsec; North is up, East to the left.}
        \label{fclus}
    \end{figure}

In the present contribution we focus our attention on compact clusters, i.e. objects that can resemble globular clusters, and old ``extended'' clusters as defined by \citet{hux}, neglecting young associations, whose brightest stars are resolved in our images.
We searched our image for compact  star clusters by visual inspection and looking for sources more extended than the PSF in source catalogues obtained with Sextractor, as described in \citet{silvr}. Our search was not intended to be exhaustive, but to single out the best candidates. The
images of the four objects we finally identified as the most reliable candidates are shown in Fig.~\ref{fclus}; their main properties are listed in Tab.~\ref{clustab}.

We re-identified Sex~B-C1 and we found two further candidates in Sex~B and one in Sex~A. We got the positions, integrated magnitudes and measured FWHM from aperture photometry that we performed with Sextractor on the g,r pair of deep  images with the best seeing. All the candidates are clearly non-stellar objects, i.e. their light profile is significantly more extended than the PSF. It is reassuring that our estimate of the 
integrated V magnitude for Sex~B-C1 is in excellent agreement with \citet{shaclus}, that report $V_0=17.90\pm 0.02$. The case of Sex~B-C1 shows that compact clusters cannot be unequivocally identified in our images (Fig.~\ref{fclus}). Indeed none of the presented candidates is particularly convincing. The hint of a diffuse halo and the relatively red color suggests that Sex~B-C2 may in fact be a roundish 
elliptical galaxy in the background \citep[see][]{pea}. The fuzzy appearance of Sex~B-C3 and Sex~A-C1 is more akin to extended clusters \citep{hux,N6822} than to classical globulars. Spectroscopic and/or high-spatial-resolution imaging follow up is clearly needed to establish the actual nature of these objects.

\end{document}